\documentclass[prd,aps,nofootinbib,superscriptaddress]{revtex4}
\usepackage{epsfig}
\usepackage{amsmath}
\usepackage{amsfonts}
\usepackage{amssymb}
\usepackage{graphicx}
\usepackage{graphics}
\usepackage{hyperref}
\usepackage{epstopdf}
\usepackage{color}
\usepackage{adjustbox,lipsum}

\definecolor{blu}{cmyk}{1,0.7,0,0.6}
\begin{document}
\title{\color{blu}Effects of Two Inert Scalar Doublets on Higgs Interactions and
Electroweak Phase Transition}
\author{Amine Ahriche}
\email{aahriche@ictp.it}
\affiliation{Department of Physics, University of Jijel, PB 98 Ouled Aissa, DZ-18000 Jijel, Algeria}
\affiliation{The Abdus Salam International Centre for Theoretical Physics, Strada Costiera
11, I-34014, Trieste, Italy}
\author{Gaber Faisel}
\email{gfaisel@hep1.phys.ntu.edu.tw}
\affiliation{Department of Physics and Center for Theoretical Sciences, National Taiwan
University, Taipei 106, Taiwan}
\affiliation{Egyptian Center for Theoretical Physics, Modern University for Information and
Technology, Cairo 11212, Egypt}
\author{Shu-Yu Ho}
\email{sho3@caltech.edu}
\affiliation{Department of Physics, California Institute of Technology, Pasadena, CA 91125, USA}
\author{Salah Nasri}
\email{snasri@uaeu.ac.ae}
\affiliation{Physics Department, UAE University, POB 17551, Al Ain, United Arab Emirates}
\author{Jusak Tandean}
\email{jtandean@yahoo.com}
\affiliation{Department of Physics and Center for Theoretical Sciences, National Taiwan
University, Taipei 106, Taiwan}

\begin{abstract}
We study some implications of the presence of two new scalar weak doublets
beyond the standard model which have zero vacuum expectation values and are
charged under an extra Abelian gauge symmetry. The additional gauge sector
does not couple directly to standard-model particles. We investigate
specifically the effects of the scalars on oblique electroweak parameters and
on the interactions of the 125 GeV Higgs boson, especially its decay modes
$h\rightarrow\gamma\gamma,\gamma Z$ and trilinear self-coupling, all of which
will be probed with improved precision in future Higgs measurements. Moreover,
we explore how the new scalars may give rise to strongly first-order
electroweak phase transition and also show its correlation with sizable
modifications to the Higgs trilinear self-coupling.

\end{abstract}

\maketitle

\section{Introduction}

The recent discovery~\cite{atlas:h,cms:h} at the Large Hadron Collider (LHC)
of a Higgs boson with mass around 125 \textrm{GeV} and other properties
consistent with the expectations of the standard model~(SM) serves as yet
another confirmation that it is a remarkably successful theory. Nevertheless,
it is widely believed that new physics beyond it is still necessary at least
to account for the compelling experimental evidence for neutrino mass and the
astronomical indications of dark matter \cite{pdg}.

Among a great many possibilities beyond the SM are those with enlarged scalar
sectors. Scenarios incorporating a second Higgs doublet are of course highly
popular in the literature \cite{thdm,Branco:2011iw}. Of late models with three
scalar weak doublets have also been gaining interest
\cite{Ivanov:2011ae,Machado:2012ed,Keus:2013hya,Fortes:2014dia,Maniatis:2014oza,
Ivanov:2014doa,Moretti:2015cwa,Keus:2014jha,Ma:2013yga}, as they can provide
dark matter (DM) candidates \cite{Keus:2014jha} and/or an important ingredient
for the mechanism that generates neutrino mass \cite{Ma:2013yga}.

Here we consider this three-scalar-doublet possibility, particularly that in
which two of the doublets possess zero vacuum expectation values (VEVs). The
theory also involves a new Abelian gauge symmetry under which these two
doublets are charged, while SM particles are not. As a consequence, the extra
scalar particles do not couple directly to a pair of exclusively SM fermions.
Because of the absence of their VEVs and couplings to SM fermion pairs, these
scalars have been termed inert in the literature~\cite{Keus:2013hya}. However,
being members of weak doublets, these scalars have interactions with SM gauge
bosons at tree level. In addition, the gauge boson associated with the new
gauge group is taken to have vanishing kinetic mixing with the hypercharge
gauge boson. Accordingly, the additional gauge sector can be regarded as dark.

With these choices, the scalar sector of the theory corresponds to one of the
three-scalar-doublet models catalogued and studied in Ref. \cite{Keus:2013hya}
in terms of all possible allowed symmetries. In the present paper, we
entertain the scenario described above and explore some implications of the
presence of the inert scalars. Specifically, we analyze constraints on them
from collider measurements on the Higgs boson and from electroweak precision
data. In addition, we look at the potential impact of the scalars on the Higgs
trilinear self-coupling, anticipating future experiments that will probe it
sufficiently well. To evaluate the coupling, we will employ the Higgs
effective potential derived at the one-loop level. Moreover, we examine how
the new particles, which we choose to have sub-TeV masses, may give rise to
strongly first-order electroweak phase transition (EWPT), which is needed for
electroweak baryogenesis to explain the baryon asymmetry of the Universe. As
it has been pointed out in the context of other models that the strength of
EWPT could be correlated with sizable modifications to the Higgs trilinear
self-coupling~\cite{KOS,Noble:2007kk,Chiang:2014hia,Fuyuto:2014yia,Curtin:2014jma}%
, our results will indicate how this may be realized in the presence of the
new doublets.

Due to their tree-level interactions with SM gauge and Higgs bosons, the
lightest of the inert scalars cannot serve as good candidates for DM, as they
annihilate into SM particles too fast and hence cannot produce enough relic
abundance. To account for DM, one needs to have a more complete theory, but we
assume that the additional ingredients responsible for explaining DM have
negligible or no impact on our scalar sector of interest, so that they do not
affect the results of this paper.\footnote{In a recently proposed scotogenic
model~\cite{Ma:2013yga}, the inert scalars participate in the mechanism to
generate light neutrino masses via one-loop interactions with new fermions
which include good DM candidates. In such a context, our results would likely
be modified.}

The plan of the paper is as follows. In the next section, we describe the
scalar Lagrangian and address some theoretical constraints on its parameters,
especially from the requirements on vacuum stability. Since the extra scalar
doublets couple to the standard Higgs and gauge bosons and include
electrically-charged members, they contribute at the one-loop level to the
Higgs decays \thinspace$h\rightarrow\gamma\gamma$\thinspace\ and
\thinspace$h\rightarrow\gamma Z$\thinspace\ which have been under intense
investigation at the LHC, the former channel having also been observed. We
determine their rates in section \ref{hcons}, where we also start our
numerical analysis by exploring the charged scalars' impact on these
processes. In section \ref{ewpconstraints}, we calculate the contributions of
the new doublets to the oblique electroweak observables $S$ and~$T$, on which
experimental information is available. Sections \ref{thc} and \ref{ewpt}
contain our treatment of the new scalars' effects on the trilinear Higgs
couplings and on the electroweak phase transition, respectively. After
deriving the relevant formulas, we perform further numerical work in these
sections. In section \ref{concl}, we discuss additional results and make our
conclusions after combining different relevant constraints. A~few appendices
contain more discussions and formulas.

\section{Scalar Sector\label{model}}

\subsection{Lagrangian\label{Lscalar}}

Compared to the SM with the Higgs doublet $\Phi$, the scalar sector is
expanded with the addition of two weak doublets, $\eta_{1}$ and $\eta_{2}$.
The theory also possesses an extra Abelian gauge symmetry, $\mathrm{U(1)}%
{}_{D}$, under which $\eta_{1,2}$ carry charges $+1$ and $-1$, respectively,
whereas SM particles are not charged. Accordingly, one can express the
renormalizable Lagrangian for the interactions of the scalars with each other
and with the standard $\mathrm{SU(2)}{}_{L}\times\mathrm{U(1)}{}_{Y}$ gauge
bosons, $W_{1,2,3}$ and $B$, as well as the $\mathrm{U(1)}{}_{D}$ gauge boson
$C$, as
\begin{equation}
\mathcal{L}=(\mathcal{D}^{\mu}\Phi)^{\dagger}\mathcal{D}_{\mu}\Phi
+(\mathcal{D}^{\mu}\eta_{1})^{\dagger}\mathcal{D}_{\mu}\eta_{1}+(\mathcal{D}%
^{\mu}\eta_{2})^{\dagger}\mathcal{D}_{\mu}\eta_{2}\;-\;\mathcal{V}%
,\label{lagrangian}%
\end{equation}
where the covariant derivative \thinspace$\mathcal{D}^{\mu}=\partial^{\mu
}+(ig/2)\tau_{j}W_{j}^{\mu}+ig_{Y}\mathcal{Q}_{Y}B^{\mu}+ig_{D}\mathcal{Q}%
_{C}C^{\mu}$\thinspace\ also contains the gauge couplings $g$, $g_{Y}$, and
$g_{D}$, Pauli matrices $\tau_{1,2,3}$, and $\mathrm{U(1)}{}_{Y,D}$ charge
operators $\mathcal{Q}_{Y,C}$, while the scalar potential is
\begin{align}
\mathcal{V}  &  =\mu_{1}^{2}\Phi^{\dagger}\Phi+\mu_{21}^{2}\eta_{1}^{\dagger
}\eta_{1}+\mu_{22}^{2}\eta_{2}^{\dagger}\eta_{2}+\tfrac{1}{2}\lambda_{1}%
(\Phi^{\dagger}\Phi)^{2}+\tfrac{1}{2}\lambda_{21}(\eta_{1}^{\dagger}\eta
_{1})^{2}+\tfrac{1}{2}\lambda_{22}(\eta_{2}^{\dagger}\eta_{2})^{2}%
\nonumber\label{potential}\\
&  ~~~+\lambda_{31}\Phi^{\dagger}\Phi\eta_{1}^{\dagger}\eta_{1}+\lambda
_{32}\Phi^{\dagger}\Phi\eta_{2}^{\dagger}\eta_{2}+\lambda_{41}\Phi^{\dagger
}\eta_{1}\eta_{1}^{\dagger}\Phi+\lambda_{42}\Phi^{\dagger}\eta_{2}\eta
_{2}^{\dagger}\Phi\nonumber\\
&  ~~~+\tfrac{1}{2}\left[  \lambda_{5}\Phi^{\dagger}\eta_{1}\Phi^{\dagger}%
\eta_{2}+\lambda_{5}^{\ast}\eta_{1}^{\dagger}\Phi\eta_{2}^{\dagger}%
\Phi\right]  +\lambda_{6}\eta_{1}^{\dagger}\eta_{1}\eta_{2}^{\dagger}\eta
_{2}+\lambda_{7}\eta_{1}^{\dagger}\eta_{2}\eta_{2}^{\dagger}\eta_{1}.
\end{align}
Thus \thinspace$\mathcal{Q}_{C}\Phi=0$\thinspace\ and \thinspace
$\mathcal{Q}_{C}\eta_{1}(\eta_{2})=+\eta_{1}(-\eta_{2})$.\thinspace\ The
parameters $\mu_{1,2a}^{2}$ and $\lambda_{1,2a,3a,4a,6,7}$ with \thinspace
$a=1,2$\thinspace\ are necessarily real because of the hermiticity of
$\mathcal{V}$, whereas $\lambda_{5}$ can be rendered real using the relative
phase between $\Phi$ and $\eta_{1,2}$. Assuming that the $\mathrm{U(1)}{}_{D}$
symmetry stays intact, after electroweak symmetry breaking we can write
\begin{equation}
\Phi=\left(
\begin{array}
[c]{c}%
0\vspace{2pt}\\
\frac{1}{\sqrt{2}}(v+h)
\end{array}
\right)  ,~~~~~~~\eta_{a}=\left(
\begin{array}
[c]{c}%
H_{a}^{+}\vspace{3pt}\\
\eta_{a}^{0}%
\end{array}
\right)  ,~~~~\sqrt{2}\;\eta_{a}^{0}=\mathrm{Re}_{{}}\eta_{a}^{0}%
+i\mathrm{Im}_{{}}\eta_{a}^{0}\;,
\end{equation}
where $h$ represents the physical Higgs boson, $v\simeq246$ \textrm{GeV} is
the vacuum expectation value (VEV) of $\Phi$, and $H_{a}^{+}$ and $\eta
_{a}^{0}$ denote, respectively, the electrically charged and neutral
components of $\eta_{a}$, which has no VEV.

From the terms in $\mathcal{V}$ that are quadratic in the fields, it is
straightforward to extract the mass eigenstates of the scalars. Thus the
masses of $h$ and $H_{1,2}^{\pm}$ at tree level are given by
\begin{equation}
\hat{m}_{h}^{2}=\mu_{1}^{2}+\tfrac{3}{2}\lambda_{1}^{~}v^{2},~~~m_{H_{a}^{\pm
}}^{2}=\mu_{2a}^{2}+\tfrac{1}{2}\lambda_{3a}v^{2}.\label{mh}%
\end{equation}
The $\lambda_{5}$ part in eq.~(\ref{potential}) causes mixing between the
electrically neutral components $\eta_{1}^{0}$ and~$\eta_{2}^{0\ast}$, which
are then related to the mass eigenstates $\chi_{1}$ and $\chi_{2}$ according
to
\begin{align}
\left(  \!%
\begin{array}
[c]{c}%
\eta_{1}^{0}\vspace{2pt}\\
\eta_{2}^{0\ast}%
\end{array}
\!\right)   &  =\left(  \!%
\begin{array}
[c]{ccc}%
c_{\theta} &  & s_{\theta}\vspace{1pt}\\
-s_{\theta} &  & c_{\theta}%
\end{array}
\!\right)  \left(  \!%
\begin{array}
[c]{c}%
\chi_{1}\vspace{1pt}\\
\chi_{2}%
\end{array}
\!\right)  ,~~~~~~~c_{\theta}=\cos\theta,~~~~s_{\theta}=\sin\theta
,\nonumber\label{theta}\\
\tan(2\theta) &  =\frac{\lambda_{5}v^{2}}{2m_{H_{2}^{\pm}}^{2}-2m_{H_{1}^{\pm
}}^{2}+\left(  \lambda_{42}-\lambda_{41}\right)  v^{2}}\;,
\end{align}
the resulting eigenmasses being given by
\begin{align}
m_{\chi_{1,2}}^{2} &  =\tfrac{1}{2}\left(  m_{H_{1}^{\pm}}^{2}+m_{H_{2}^{\pm}%
}^{2}\right)  +\tfrac{1}{4}\left(  \lambda_{41}+\lambda_{42}\right)
v^{2}\nonumber\\
&  ~~~\mp\tfrac{1}{2}\sqrt{\left[  m_{H_{2}^{\pm}}^{2}-m_{H_{1}^{\pm}}%
^{2}+\tfrac{1}{2}\left(  \lambda_{42}-\lambda_{41}\right)  v^{2}\right]
^{2}+\tfrac{1}{4}\lambda_{5}^{2}v^{4}}\;.\label{mx}%
\end{align}
Hence the $\mathrm{U(1)}{}_{D}$ charges of $\chi_{1,2}$ are the same as
(opposite in sign to) that of $\eta_{1}$ ($\eta_{2}$) and $m_{\chi_{1}}\leq
m_{\chi_{2}}$.

Alternatively, instead of $\chi_{a}$, one can choose to deal with their real
and imaginary parts,
\begin{equation}
\mathcal{S}_{a}=\sqrt{2}\;\mathrm{Re}_{ }\chi_{a} ,~~~~\mathcal{P}_{a}%
=\sqrt{2}\;\mathrm{Im}_{ }\chi_{a} ,\label{SaPa}%
\end{equation}
which are $CP$-even and $CP$-odd states, respectively, and share mass,
$m_{\mathcal{S}_{a}}=m_{\mathcal{P}_{a}}=m_{\chi_{a}}$. From eq.~(\ref{theta}%
), one then has in matrix form
\begin{equation}
\left(
\begin{array}
[c]{c}%
\mathrm{Re}_{ }\eta_{1}^{0}\\
\mathrm{Re}_{ }\eta_{2}^{0}\\
\mathrm{Im}_{ }\eta_{1}^{0}\\
\mathrm{Im}_{ }\eta_{2}^{0}%
\end{array}
\right)  =\left(
\begin{array}
[c]{ccccccc}%
c_{\theta} &  & s_{\theta} &  & 0 &  & 0\\
-s_{\theta} &  & c_{\theta} &  & 0 &  & 0\\
0 &  & 0 &  & c_{\theta} &  & s_{\theta}\\
0 &  & 0 &  & s_{\theta} &  & -c_{\theta}%
\end{array}
\right)  \left(
\begin{array}
[c]{c}%
\mathcal{S}_{1}\\
\mathcal{S}_{2}\\
\mathcal{P}_{1}\\
\mathcal{P}_{2}%
\end{array}
\right)  ,\label{mix}%
\end{equation}
where the mixing matrix is orthogonal.

Later on, to simplify the analysis, we will concentrate on the scenario in
which $\lambda_{5}$ is negligible compared to the other $\lambda$'s in
$\mathcal{V}$. In that case, as eq.~(\ref{theta}) indicates, the $\eta_{1}%
^{0}$-$\eta_{2}^{0\ast}$ mixing is small, $\theta\ll1$, provided that
$\lambda_{5}v^{2}\ll2m_{H_{1}^{\pm}}^{2}-2m_{H_{2}^{\pm}}^{2}+\left(
\lambda_{41}-\lambda_{42}\right)  v^{2}$.\ Furthermore, one can see from
eq.~(\ref{mx}) that at the same time $\chi_{1}$ and $\chi_{2}$ can be close in
mass if $\tfrac{1}{2}\lambda_{5}v^{2}\ll\left\vert m_{H_{1}^{\pm}}%
^{2}-m_{H_{2}^{\pm}}^{2}+\tfrac{1}{2}\left(  \lambda_{41}-\lambda_{42}\right)
v^{2}\right\vert \ll m_{\chi_{1}}^{2}$.

\subsection{Theoretical Constraints\label{tconstraints}}

The parameters of the scalar potential are subject to a number of theoretical
constraints. The stability of the vacuum implies that $\mathcal{V}$ must be
bounded from below. As shown in appendix \ref{app:vacstab}, with $\lambda_{5}$
being negligible, this entails that for $a=1,2$
\begin{gather}
\lambda_{1}>0 ,~~~\lambda_{2a}>0 ,~~~\lambda_{3a}+\lambda_{4a}^{0}
+\sqrt{\lambda_{1}\lambda_{2a}} >0 ,~~~\lambda_{6}+\lambda_{7}^{0}
+\sqrt{\lambda_{21}\lambda_{22}} >0 ,\nonumber\\
\sqrt{\lambda_{1}\lambda_{21}\lambda_{22}}+\sqrt{\lambda_{1}} \left(
\lambda_{6}+\lambda_{7}^{0}\right)  +\sqrt{\lambda_{21}}\left(  \lambda
_{32}+\lambda_{42}^{0}\right)  +\sqrt{\lambda_{22}}\left(  \lambda
_{31}+\lambda_{41}^{0}\right) \nonumber\\
+\left[  2\left(  \sqrt{\lambda_{1}\lambda_{21}}+\lambda_{31}+\lambda_{41}%
^{0}\right)  \left(  \sqrt{\lambda_{1}\lambda_{22}}+\lambda_{32}+\lambda
_{42}^{0}\right)  \left(  \sqrt{\lambda_{21}\lambda_{22}}+\lambda_{6}%
+\lambda_{7}^{0}\right)  \right]  ^{1/2}> 0 ,\label{vacs}%
\end{gather}
where $\lambda_{x}^{0}\equiv\mathrm{Min}(0,\lambda_{x})$.

The $\mu^{2}$ and $\lambda$ parameters in $\mathcal{V}$ also need to have such
values that its minimum with the VEV of $\Phi$ ($\eta_{a}$) being nonzero
(zero) is global. This is already guaranteed~\cite{Keus:2013hya} by the
positivity of the mass eigenvalues in eqs. (\ref{mh}) and (\ref{mx}).

In addition, the perturbativity of the theory implies that the magnitudes of
the $\lambda$ parameters need to be capped. Thus, in numerical work our
choices for their ranges, to be specified later on, will meet the general
requirement $|\lambda_{x}|<8\pi$, in analogy to that in the two-Higgs-doublet
case~\cite{Kanemura:1999xf}.

\section{Restrictions from Collider Data\label{hcons}}

The kinetic portion of the Lagrangian in eq.~(\ref{lagrangian}) contains the
interactions of the new scalars with the photon and weak bosons,
\begin{align}
\mathcal{L}  &  \supset iH_{a}^{+}%
\mbox{\footnotesize$\stackrel{\scriptscriptstyle\leftrightarrow}{\partial}$}{}%
^{\mu}H_{a}^{-}\left(  eA_{\mu}-g_{L}Z_{\mu}\right)  +H_{a}^{+}H_{a}%
^{-}\left(  eA-g_{L}Z\right)  ^{2}\nonumber\label{Lint}\\
&  ~~~+\frac{ig}{2c_{\mathrm{w}}}\left[  c_{2\theta}\left(  \chi_{1}^{\ast}
\mbox{\footnotesize$\stackrel{\scriptscriptstyle\leftrightarrow}{\partial}$}{}%
^{\mu}\chi_{1}-\chi_{2}^{\ast}
\mbox{\footnotesize$\stackrel{\scriptscriptstyle\leftrightarrow}{\partial}$}{}%
^{\mu}\chi_{2}\right)  +s_{2\theta}\left(  \chi_{1}^{\ast}%
\mbox{\footnotesize$\stackrel{\scriptscriptstyle\leftrightarrow}{\partial}$}{}%
^{\mu}\chi_{2}+\chi_{2}^{\ast}
\mbox{\footnotesize$\stackrel{\scriptscriptstyle\leftrightarrow}{\partial}$}{}%
^{\mu}\chi_{1}\right)  \right]  Z_{\mu}+\frac{g^{2}}{4c_{\mathrm{w}}^{2}}%
\chi_{a}^{\ast}\chi_{a}Z^{2}\nonumber\\
&  ~~~+\frac{ig}{\sqrt{2}}\left\{  \left[  c_{\theta}\left(  H_{1}%
^{+}%
\mbox{\footnotesize$\stackrel{\scriptscriptstyle\leftrightarrow}{\partial}$}{}%
^{\mu}\chi_{1}^{\ast}+H_{2}^{+}
\mbox{\footnotesize$\stackrel{\scriptscriptstyle\leftrightarrow}{\partial}$}{}%
^{\mu}\chi_{2}\right)  +s_{\theta}\left(  H_{1}^{+}%
\mbox{\footnotesize$\stackrel{\scriptscriptstyle\leftrightarrow}{\partial}$}{}%
^{\mu}\chi_{2}^{\ast}-H_{2}^{+}%
\mbox{\footnotesize$\stackrel{\scriptscriptstyle\leftrightarrow}{\partial}$}{}%
^{\mu}\chi_{1}\right)  \right]  W_{\mu}^{-}-\mathrm{H.c.}\right\} \nonumber\\
&  ~~~+\frac{g^{2}}{2}\left(  H_{a}^{+}H_{a}^{-}+\chi_{a}^{\ast}\chi
_{a}\right)  W^{+\mu}W_{\mu}^{-}\;,
\end{align}
where summation over $a=1,2$\ is implicit,
\begin{equation}
X\raisebox{1pt}{\footnotesize$\stackrel{\scriptscriptstyle\leftrightarrow}{\partial}$}{}%
^{\mu}Y=X\partial^{\mu}Y-Y\partial^{\mu}X\;,~~~~~~~g_{L}=\frac{g}%
{2c_{\mathrm{w}}}\left(  2s_{\mathrm{w}}^{2}-1\right)  ,
\end{equation}
$c_{\mathrm{w}}=\cos\theta_{\mathrm{w}}=(1-s_{\mathrm{w}}^{2}){}^{1/2}$, with
$\theta_{\mathrm{w}}$ being the usual Weinberg angle, $c_{2\theta}%
=\cos(2\theta)$, and~$s_{2\theta}=\sin(2\theta)$. One can alternatively write
eq.~(\ref{Lint}) in terms of the real and imaginary components $\mathcal{S}%
_{a}$ and $\mathcal{P}_{a}$ of~$\chi_{a}$, which becomes more lengthy and is
relegated to appendix \ref{interactions}.

We now see that data from past colliders can lead to some constraints on the
masses of the new scalars. Based on eq.~(\ref{Lint}), we may infer from the
experimental widths of the $W$ and $Z$ bosons and the absence so far of
evidence for nonstandard particles in their decay modes that for $a,b=1,2$
\begin{equation}
m_{H_{a}^{\pm}}+m_{\chi_{b}}>m_{W},~~~~~~~2m_{H_{a}^{\pm}}>m_{Z}%
,~~~~~~~m_{\chi_{a}}+m_{\chi_{b}}>m_{Z}.\label{WZ}%
\end{equation}
The null results of direct searches for new particles at $e^{+}e^{-}$
colliders also imply lower limits on these masses, especially those of the
charged scalars.\footnote{A recent investigation~\cite{Ho:2013spa} concerning
the effects of the corresponding particles in the simplest scotogenic
model~\cite{Ma:2006km} on the relevant processes measured at LEP II suggests
that such charged scalars may face significant constraints if their masses are
below 100~\textrm{GeV}.} For these reasons, in our numerical work we will
generally consider the mass regions $m_{\chi_{a}}\geq50$ \textrm{GeV} and
$m_{H_{a}}\geq100$ \textrm{GeV}.

In addition to the requirements in the preceding paragraph and the vacuum
stability conditions in eq.~(\ref{vacs}), when selecting the inert scalars'
parameters we take into account also the Higgs mass which will be estimated at
the one-loop level in section~\ref{thc} and then limited to $m_{h}%
=(125.1\pm0.1)$ \textrm{GeV}, well within the ranges of the newest
measurements~\cite{mhx,CMS:2014ega}. More specifically, we will therefore make
the parameter choices
\begin{align}
0  &  <\lambda_{2a},\left\vert \lambda_{3a}\right\vert ,\left\vert
\lambda_{4a}\right\vert ,\left\vert \lambda_{6}\right\vert ,\left\vert
\lambda_{7}\right\vert <3,~~~~~\left\vert \mu_{2a}^{2}\right\vert <\left(
800~\mathrm{GeV}\right)  ^{2},\nonumber\\
\left\vert \lambda_{5}\right\vert  &  <0.01 \mathrm{Min}\left(  \lambda
_{2a},\left\vert \lambda_{3a}\right\vert ,\left\vert \lambda_{4a}\right\vert
,\left\vert \lambda_{6}\right\vert ,\left\vert \lambda_{7}\right\vert \right)
.\label{PAR}%
\end{align}

The recently discovered Higgs boson may offer a window into physics beyond
the~SM. The presence of new particles can give rise to modifications to the
standard decay modes of the Higgs and/or cause it to undergo exotic
decays~\cite{Curtin:2013fra}. As data from the LHC will continue to accumulate
with improving precision, they may uncover clues of new physics in the Higgs
couplings or, otherwise, yield growing constraints on various models. Here we
address some of the potential implications for our scenario of interest.
Especially, the existing experimental information on the possible Higgs decay
into invisible/nonstandard final states
\cite{Falkowski:2013dza,Giardino:2013bma,Ellis:2013lra,Belanger:2013xza,Cheung:2014noa}
and on the observed $h\to\gamma\gamma$ mode \cite{Aad:2014eha,CMS:2014ega} can
supply further restrictions on the inert scalars.

The Higgs boson couples to a pair of them according to
\begin{align}
\mathcal{L}  &  \supset\frac{2h}{v}\left[  \left(  \mu_{21}^{2}-m_{H_{1}}%
^{2}\right)  H_{1}^{+}H_{1}^{-}+\left(  \mu_{22}^{2}-m_{H_{2}}^{2}\right)
H_{2}^{+}H_{2}^{-}\right. \nonumber\\
&  ~~~~+\left.  \left(  c_{\theta}^{2}\mu_{21}^{2}+s_{\theta}^{2}\mu_{22}%
^{2}-m_{\chi_{1}}^{2}\right)  \chi_{1}^{\ast}\chi_{1}+\left(  c_{\theta}%
^{2}\mu_{22}^{2}+s_{\theta}^{2}\mu_{21}^{2}-m_{\chi_{2}}^{2}\right)  \chi
_{2}^{\ast}\chi_{2}\right. \nonumber\\
&  ~~~~+\left.  c_{\theta}s_{\theta}\left(  \mu_{21}^{2}-\mu_{22}^{2}\right)
\left(  \chi_{1}^{\ast}\chi_{2}+\chi_{2}^{\ast}\chi_{1}\right)  \right]
,\label{h2xx}%
\end{align}
from the $\mathcal{V}$ part of eq.~(\ref{lagrangian}). In view of the mass
choices made above, it follows that the decay modes $h\rightarrow\chi
_{a}^{\ast}\chi_{b}$, if kinematically allowed, contribute at tree level to
the total width of the Higgs boson and are the leading channels into
nonstandard final states in the model. Their rates have the form
\begin{equation}
\Gamma\left(  h\rightarrow\chi_{a}^{\ast}\chi_{b}\right)  =\frac{\left\vert
C_{\chi_{a}^{\ast}\chi_{b}}\right\vert ^{2}}{4\pi m_{h}^{3}v^{2}}\sqrt{\left(
m_{h}^{2}-m_{\chi_{a}}^{2}-m_{\chi_{b}}^{2}\right)  ^{2}-4m_{\chi_{a}}%
^{2}m_{\chi_{b}}^{2}},
\end{equation}
where
\begin{align}
C_{\chi_{1}^{\ast}\chi_{1}}  &  =c_{\theta}^{2}\mu_{21}^{2}+s_{\theta}^{2}%
\mu_{22}^{2}-m_{\chi_{1}}^{2},~~~~~~~C_{\chi_{2}^{\ast}\chi_{2}}=c_{\theta
}^{2}\mu_{22}^{2}+s_{\theta}^{2}\mu_{21}^{2}-m_{\chi_{2}}^{2},\nonumber\\
C_{\chi_{1}^{\ast}\chi_{2}}  &  =C_{\chi_{2}^{\ast}\chi_{1}}=c_{\theta
}s_{\theta}\left(  \mu_{21}^{2}-\mu_{22}^{2}\right)  .
\end{align}
The combined branching ratio of these decays is
\begin{equation}
\mathcal{B}\left(  h\rightarrow\chi^{\ast}\chi^{\prime}\right)  =\frac
{\sum_{a,b}\Gamma\left(  h\rightarrow\chi_{a}^{\ast}\chi_{b}\right)  }%
{\Gamma_{h}^{\mathrm{SM}}+\sum_{a,b}\Gamma\left(  h\rightarrow\chi_{a}^{\ast
}\chi_{b}\right)  },
\end{equation}
where $\Gamma_{h}^{\mathrm{SM}}$ is the SM Higgs total width and only channels
satisfying $m_{\chi_{a}}+m_{\chi_{b}}<m_{h}$ contribute to the sums.
Numerically, we adopt $\Gamma_{h}^{\mathrm{SM}}=4.08\;$\textrm{MeV}%
~\cite{lhctwiki} corresponding to $m_{h}=125.1$ \textrm{GeV}. If these
channels are open, we will require $\mathcal{B}\bigl(h\rightarrow\chi
_{a}^{\ast}\chi_{b}\bigr)<0.19$,\ based on the latest analysis of the Higgs
data \cite{Falkowski:2013dza,Giardino:2013bma,Ellis:2013lra,
Belanger:2013xza,Cheung:2014noa}.

\begin{figure}[h]
\begin{center}
\includegraphics[width = 0.48\textwidth]{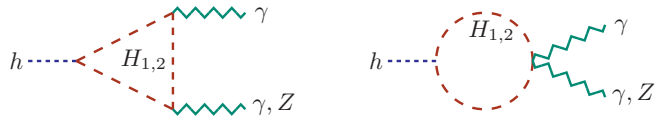}
\end{center}
\par
\vspace{-2ex}\caption{Feynman diagrams for the contributions of the new
charged scalars $H_{1,2}^{\pm}$ to the Higgs boson decays $h\to\gamma\gamma$,
and $h\to\gamma Z$. The triangle diagram with the gauge boson legs
interchanged is not shown.}%
\label{h->gg}%
\end{figure}

The potential impact of the inert scalars can also be realized through loop
diagrams. Of much interest are their contributions to the standard decay
channels $h\rightarrow\gamma\gamma$ and $h\rightarrow\gamma Z$,\ which are
already under investigation at the LHC. In the SM, they arise mainly from
top-quark- and $W$-boson-loop diagrams. These modes receive additional
contributions arising from the $H_{1,2}^{\pm}$-loop diagrams drawn in
figure~\ref{h->gg}, with vertices from eqs.~(\ref{Lint}) and (\ref{h2xx}%
).\footnote{At the one-loop level, the charged (charged and neutral) inert
scalars also induce $h\rightarrow\gamma C$ ($h\rightarrow ZC,CC$) involving
the massless dark gauge boson $C$. These decay modes may be challenging to
detect with $C$ being invisible, as their rates are expected to be roughly of
similar order to those of the $\gamma\gamma$ and $\gamma Z$ channels.} Their
decay rates are readily obtainable from those in the case of only one inert
doublet~\cite{Ho:2013hia}. Thus we get
\begin{align}
\Gamma(h\rightarrow\gamma\gamma) &  =\frac{\alpha^{2}G_{F}m_{h}^{3}}%
{128\sqrt{2}\pi^{3}}\left\vert \frac{4}{3}A_{1/2}^{\gamma\gamma}(\kappa
_{t})+A_{1}^{\gamma\gamma}(\kappa_{W})+\sum_{a=1}^{2}\frac{m_{H_{a}^{\pm}}%
^{2}-\mu_{2a}^{2}}{m_{H_{a}}^{2}}A_{0}^{\gamma\gamma}\left(  \kappa
_{H_{a}^{\pm}}\right)  \right\vert ^{2},~~~~\label{h2gg}\\
\Gamma(h\rightarrow\gamma Z) &  =\frac{\alpha G_{F}^{2}m_{W}^{2}\left(
m_{h}^{2}-m_{Z}^{2}\right)  ^{3}}{64\pi^{4}m_{h}^{3}}\left\vert \frac
{6-16s_{\mathrm{w}}^{2}}{3c_{\mathrm{w}}}A_{1/2}^{\gamma Z}(\kappa_{t}%
,\zeta_{t})+c_{\mathrm{w}}A_{1}^{\gamma Z}(\kappa_{W},\zeta_{W})\right.
\nonumber\label{h2gz}\\
&  \hspace{20ex}\left.  -~\frac{1-2s_{\mathrm{w}}^{2}}{c_{\mathrm{w}}}%
\sum_{a=1}^{2}\frac{m_{H_{a}^{\pm}}^{2}-\mu_{2a}^{2}}{m_{H_{a}}^{2}}%
A_{0}^{\gamma Z}\left(  \kappa_{H_{a}^{\pm}},\zeta_{H_{a}^{\pm}}\right)
\right\vert ^{2},
\end{align}
where $\alpha=g^{2}s_{\mathrm{w}}^{2}/(4\pi)$ is the fine-structure constant,
the expressions for the form factors $A_{0,1/2,1}^{\gamma\gamma,\gamma Z}$ are
available from ref.~\cite{Chen:2013vi}, the $A_{0}^{\gamma\gamma,\gamma Z}$
terms originate exclusively from the $H_{1,2}^{\pm}$ diagrams, $\kappa
_{X}=4m_{X}^{2}/m_{h}^{2}$, \ and~$\zeta_{X}=4m_{X}^{2}/m_{Z}^{2}$.

We can already test the new contributions to $h\rightarrow\gamma\gamma
$,\ which has been observed at the LHC, unlike the $\gamma Z$ channel. For the
$\gamma\gamma$ signal strengths, the ATLAS and CMS Collaborations
measured~$\sigma/\sigma_{\mathrm{SM}}=1.17\pm0.27$ \cite{Aad:2014eha} and
$1.13\pm0.24$ \cite{CMS:2014ega}, respectively. These numbers need to be
respected by the ratio of $\Gamma(h\rightarrow\gamma\gamma)$ to its SM value,
\begin{equation}
\mathcal{R}_{\gamma\gamma} = \frac{\Gamma(h\rightarrow\gamma\gamma)}%
{\Gamma(h\rightarrow\gamma\gamma)_{\mathrm{SM}}} \;.\label{Rgg}%
\end{equation}
Its $\gamma Z$ counterpart,
\begin{equation}
\mathcal{R}_{\gamma Z} = \frac{\Gamma(h\rightarrow\gamma Z)}{\Gamma
(h\rightarrow\gamma Z)_{\mathrm{SM}}} \;,\label{RYZ}%
\end{equation}
will be probed by future experiments.

\begin{figure}[t]
\begin{center}
\includegraphics[width = 0.48\textwidth]{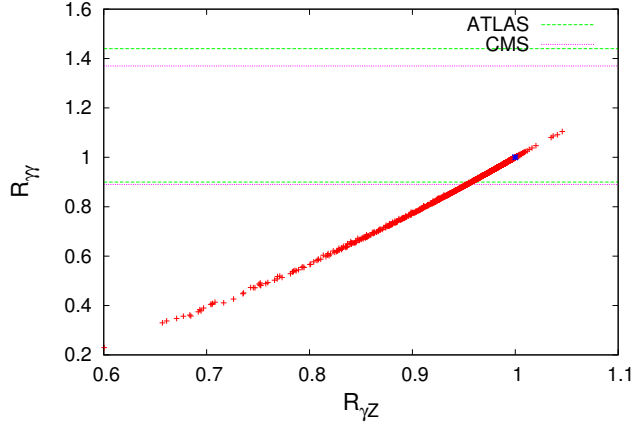}
\end{center}
\caption{The effects of the new charged scalars $H_{1,2}^{\pm}$ on the ratios
of the rates of Higgs decay channels $h\rightarrow\gamma\gamma$ and
$h\rightarrow\gamma Z$ to their respective SM values for 5000 benchmark points
as described in the text. The blue point marks the SM value. The region
between the green (magenta) horizontal lines represents the one-sigma range of
the ATLAS (CMS) data \cite{Aad:2014eha,CMS:2014ega}.}%
\label{hgz}%
\end{figure}

To illustrate the effects of the inert scalars on $\mathcal{R}_{\gamma\gamma}
$ and $\mathcal{R}_{\gamma Z}$, and possible (anti)correlation between them,
we display in figure \ref{hgz} the distribution of 5000 benchmark points on
the $(\mathcal{R}_{\gamma Z},\mathcal{R}_{\gamma\gamma})$ plane which satisfy
the vacuum stability requirements in eq.~(\ref{vacs}), the constraints from
$W$ and $Z$ decays in eq.~(\ref{WZ}), and the parameter limitations in
eq.~(\ref{PAR}). We notice that many of the $\mathcal{R}_{\gamma\gamma}$
values are close to 1 and within the allowed ranges from ATLAS and CMS. The
plot also reveals that for the $\mathcal{R}_{\gamma\gamma} $ points compatible
with the LHC data the values of $\Gamma(h\rightarrow\gamma Z)$ do not differ
from its SM value by more than 10\% or so. Furthermore, there is a positive
correlation between $\mathcal{R}_{\gamma\gamma}$ and $\mathcal{R}_{\gamma Z}$,
which is much like the situations in a different recent model with two inert
doublets~\cite{Fortes:2014dia} and in the case of only one inert
doublet~\cite{Ho:2013hia,Swiezewska:2012eh,Banik:2014cfa}. This can be checked
experimentally when the $\gamma Z$ mode is observed in the future.

\section{Electroweak Precision Tests\label{ewpconstraints}}

The interactions of the new doublets with the SM gauge bosons described by
eq.~(\ref{Lint}) bring about modifications, $\Delta S$ and $\Delta T$, to the
so-called oblique electroweak parameters $S$ and $T$ which encode the effects
of new physics not directly coupled to SM fermions \cite{Peskin:1991sw}. At
the one-loop level \cite{Peskin:1991sw,pdg}
\begin{align}
\frac{\alpha\Delta S}{4c_{\mathrm{w}}^{2}s_{\mathrm{w}}^{2}}  &  =\frac
{A_{ZZ}\left(  m_{Z}^{2}\right)  -A_{ZZ}(0)}{m_{Z}^{2}}-A_{\gamma\gamma
}^{\prime}(0)-\frac{c_{\mathrm{w}}^{2}-s_{\mathrm{w}}^{2}}{c_{\mathrm{w}%
}s_{\mathrm{w}}}A_{\gamma Z}^{\prime}(0),\nonumber\label{ST}\\
\alpha\Delta T  &  =\frac{A_{WW}(0)}{m_{W}^{2}}-\frac{A_{ZZ}(0)}{m_{Z}^{2}},
\end{align}
where the functions $A_{XY}\left(  q^{2}\right)  $ can be extracted from the
vacuum polarization tensors $\Pi_{XY}^{\mu\nu}\left(  q^{2}\right)
=A_{XY}\left(  q^{2}\right)  g^{\mu\nu}+[q^{\mu}q^{\nu}\mathrm{\;terms}%
]$\thinspace\ of the SM gauge bosons due to the new scalars' loop
contributions, and $A_{XY}^{\prime}(0)=[dA_{XY}\left(  q^{2}\right)
/dq^{2}]_{q^{2}=0}$. In our numerical analysis below, we will impose%
\begin{equation}
\Delta S=0.05\pm0.11,~~~~~\Delta T=0.09\pm0.13,
\end{equation}
which are based on the results of a recent fit \cite{Baak:2014ora} to
electroweak precision data for a~Higgs mass $m_{h}=125$ \textrm{GeV}.

\begin{figure}[h]
\begin{center}
\includegraphics[width = 0.48\textwidth]{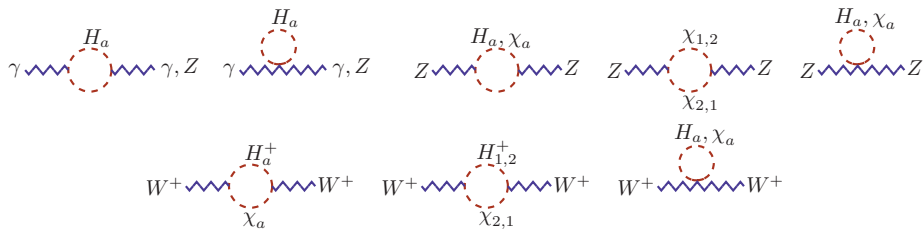}
\end{center}
\caption{Feynman diagrams for the contributions of the inert scalar doublets
to the oblique electroweak parameters $\Delta S$ and $\Delta T$.}%
\label{PP}%
\end{figure}

The contributions of the inert scalars to $\Delta S$ and $\Delta T$ arise from
the diagrams depicted in figure~\ref{PP}. After evaluating them, we arrive
at\footnote{Their counterparts in the case of only one inert scalar doublet
were computed in Ref. \cite{Barbieri:2006dq}.}%
\begin{align}
\Delta S &  =\frac{1}{6\pi}\left[  \ln\frac{m_{\chi_{1}}m_{\chi_{2}}}%
{m_{H_{1}^{\pm}}m_{H_{2}^{\pm}}}+s_{2\theta}^{2}\frac{22m_{\chi_{1}}%
^{2}m_{\chi_{2}}^{2}-5m_{\chi_{1}}^{4}-5m_{\chi_{2}}^{4}}{6\left(  m_{\chi
_{1}}^{2}-m_{\chi_{2}}^{2}\right)  ^{2}}\right. \nonumber\\
&  \hspace{8ex}\left.  +\;s_{2\theta}^{2}\frac{\left(  m_{\chi_{1}}%
^{2}+m_{\chi_{2}}^{2}\right)  \left(  m_{\chi_{1}}^{4}-4m_{\chi_{1}}%
^{2}m_{\chi_{2}}^{2}+m_{\chi_{2}}^{4}\right)  }{\left(  m_{\chi_{1}}%
^{2}-m_{\chi_{2}}^{2}\right)  ^{3}}\ln\frac{m_{\chi_{1}}}{m_{\chi_{2}}%
}\right]  ,\\
\Delta T &  =\frac{1}{8\alpha_{\;}\!\pi^{2}v^{2}}\left[  c_{\theta}%
^{2}\mathcal{F}\left(  m_{H_{1}^{\pm}},m_{\chi_{1}}\right)  +c_{\theta}%
^{2}\mathcal{F}\left(  m_{H_{2}^{\pm}},m_{\chi_{2}}\right)  +s_{\theta}%
^{2}\mathcal{F}\left(  m_{H_{1}^{\pm}},m_{\chi_{2}}\right)  \right.
\nonumber\\
&  \hspace{12ex}\left.  +s_{\theta}^{2}\mathcal{F}\left(  m_{H_{2}^{\pm}%
},m_{\chi_{1}}\right)  -4c_{\theta}^{2}s_{\theta}^{2}\mathcal{F}\left(
m_{\chi_{1}},m_{\chi_{2}}\right)  \right]  ,
\end{align}
where
\begin{equation}
\mathcal{F}(m,n)=\frac{m^{2}+n^{2}}{2}-\frac{m^{2}n^{2}}{m^{2}-n^{2}}\ln
\frac{m^{2}}{n^{2}}.
\end{equation}

\begin{figure}[t]
\includegraphics[width = 0.48\textwidth]{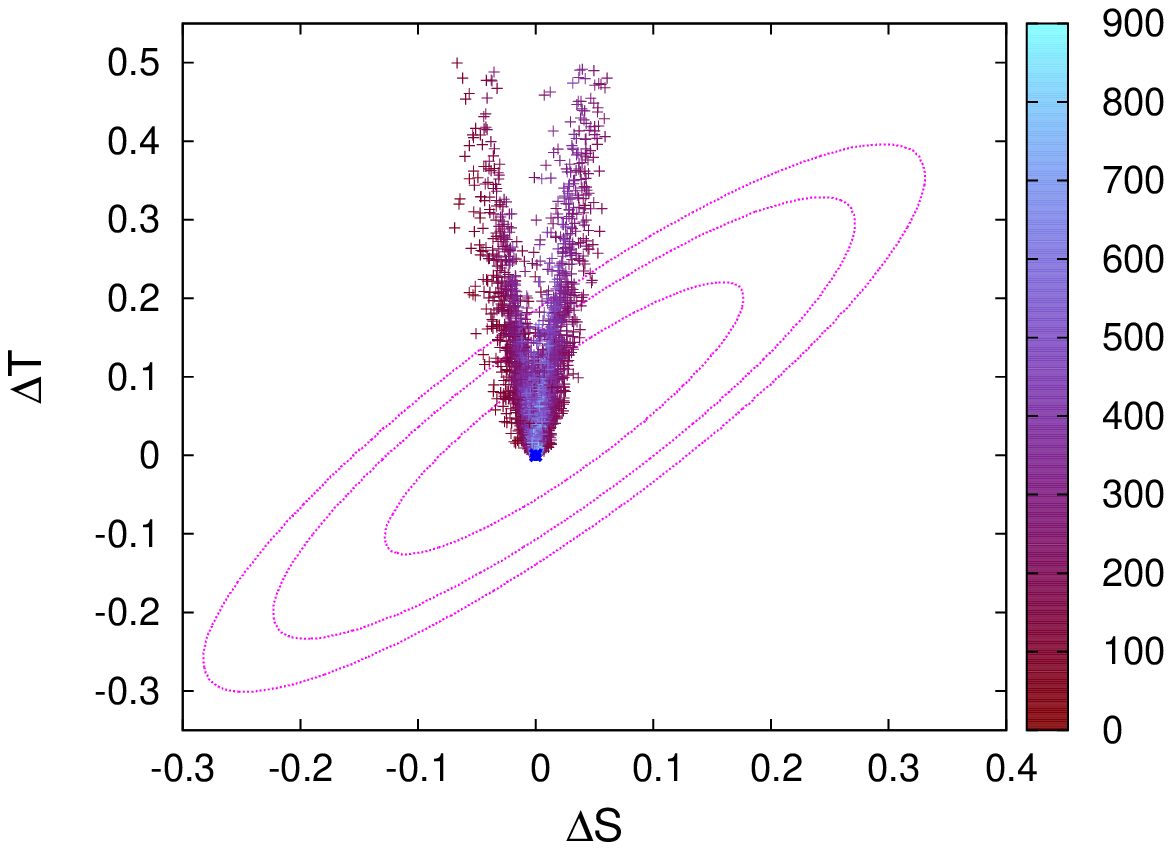}~\includegraphics[width =
0.48\textwidth]{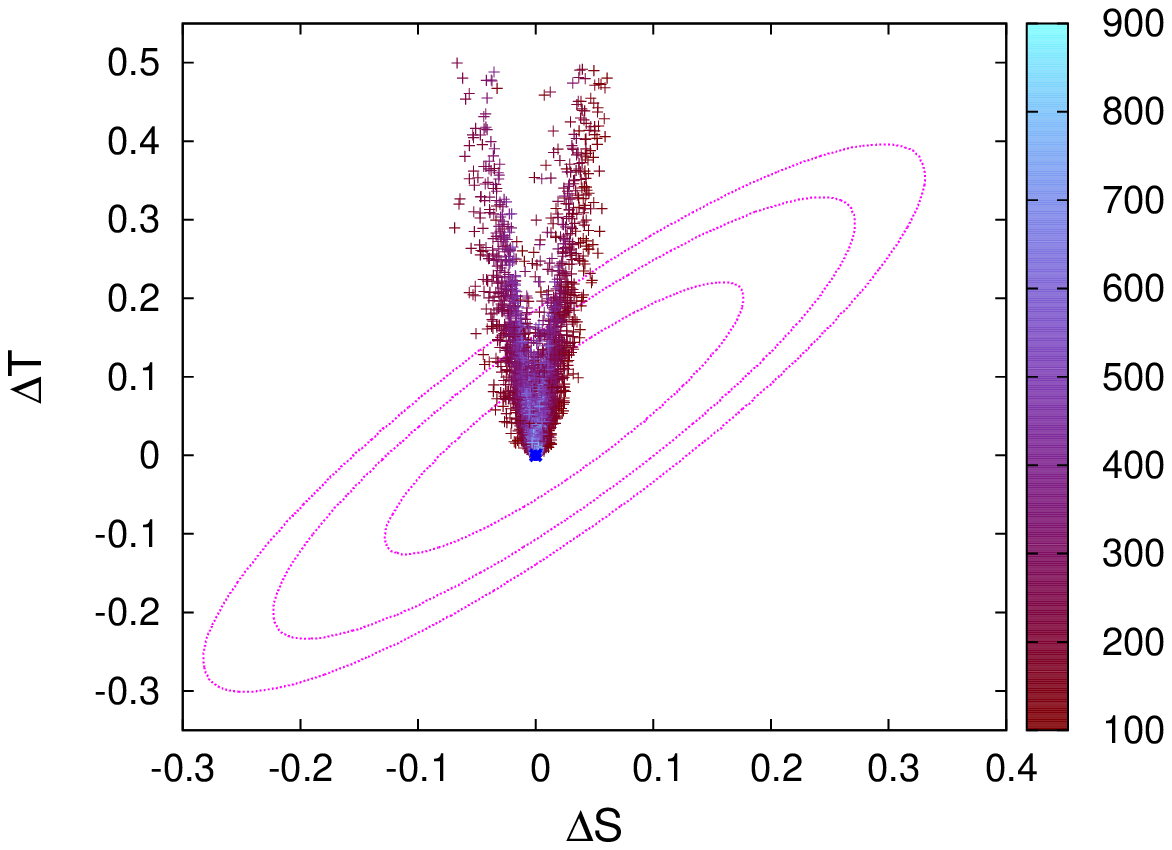}\vspace{-3pt}\caption{The contributions of the
inert scalar doublets to the oblique electroweak parameters $\Delta S$ and
$\Delta T$ for the 5000 benchmarks used previously. On the left panel, the
palette belongs to the lighter neutral inert scalar's mass, $m_{\chi_{1}}$, in
GeV. On the right panel, the palette belongs to the lighter charged scalar
mass, $m_{H_{1}}$, in GeV. The different contours represent 68\%, 95\%, and
99\% confidence level, respectively. The blue point at (0,0) marks the SM
value.}%
\label{ob}%
\end{figure}

In figure \ref{ob}, we present the distribution on the $(\Delta S,\Delta T)$
plane of the inert scalars' contributions for the 5000 benchmarks employed
previously for figure \ref{hgz}. Evidently, it is possible for the masses of
the charged scalars to be as small as 100\thinspace\textrm{GeV} and still be
compatible with the electroweak precision measurements. However, we find that
the lighter one of the inert neutral scalars, $\chi_{1}$, must be heavier than
about 90\thinspace\textrm{GeV}, which is a stronger condition than that
inferred from the LEP constraint on the invisible width of the $Z$ boson. This
also makes the bound from the data on the Higgs invisible/nonstandard decay irrelevant.

\section{Higgs Trilinear Coupling\label{thc}}

Since the new scalars couple directly to the Higgs boson, their presence can
cause its trilinear self-coupling, $\lambda_{hhh}$, to shift from its SM
prediction. Such a modification could translate into detectable collider
signatures, especially at a future $e^{+}e^{-}$ machine such as the
International Linear Collider \cite{ILCnew} where the coupling can be measured
with 20\% precision or better at a center-of-mass energy $\sqrt{s}=500$
\textrm{GeV} if the integrated luminosity is $500\;\mathrm{fb}^{-1}$.

To derive the formula for the mass-dimension Higgs trilinear self-coupling in
the presence of extra heavy particles, we follow the steps taken in
ref.~\cite{AAN}. It is just the third derivative of the Higgs effective
potential, namely
\begin{equation}
\lambda_{hhh}=\left.  \frac{\partial^{3}}{\partial\varphi^{3}}%
V_{\mathrm{\mathrm{eff}}}^{T=0}(\varphi)\right\vert _{\varphi=v},\label{l3h}%
\end{equation}
where $\varphi$ is the classical Higgs field and $V_{\mathrm{\mathrm{eff}}%
}^{T=0}(\varphi)$ is the potential evaluated at temperature $T=0$. We estimate
the potential at the one-loop level in the so-called $\overline{\mathrm{DR}}%
{}^{\prime}$ scheme \cite{NaLa,Martin:2001vx} where it has the form
\begin{equation}
V_{\mathrm{eff}}^{T=0}(\varphi)=\frac{\mu_{1}^{2}}{2}\varphi^{2}+\frac
{\lambda_{1}}{8}\varphi^{4}+\sum_{i}n_{i}\frac{\left(  m_{i}^{2}%
(\varphi)\right)  ^{2}}{64\pi^{2}}\left(  \ln\frac{m_{i}^{2}(\varphi)}%
{\Lambda^{2}}-\frac{3}{2}\right)  .\label{Veff}%
\end{equation}
In the sum above, the index $i$ runs over all the contributing particles,
$n_{i}$ stands for the number of internal degrees of freedom of the $i$th
particle, with a minus sign added if it is a fermion, $m_{i}^{2}(\varphi)$ is
its field-dependent squared mass, and $\Lambda$ is the renormalization scale
which we choose to be the Higgs mass, $\Lambda=125.1$~\textrm{GeV}. More
explicitly, $n_{h}=1$, $n_{\mathcal{G}}=n_{Z}=n_{\gamma}=3$, $n_{W}=6$,
$n_{t}=-12$, and $n_{\chi_{a}}=n_{H_{a}^{\pm}}=2$, where $\mathcal{G}$ refers
to the Goldstone bosons. We have collected the formulas for the various
relevant $m_{i}^{2}(\varphi)$ in appendix \ref{masses}.

At tree level we have $\mu_{1}^{2}=-\lambda_{1}v^{2}/2\equiv\hat{\mu}_{1}^{2}
$, but it receives the one-loop correction%
\begin{equation}
\delta\mu_{1}^{2}=-\frac{1}{32\pi^{2}v}\sum_{i}\left.  n_{i}m_{i}^{2}\dot
{m}_{i}^{2}\left(  \ln\frac{m_{i}^{2}}{\Lambda^{2}}-1\right)  \right\vert
_{\varphi=v},\label{dmu}%
\end{equation}
which follows from $\partial V_{\mathrm{eff}}^{T=0}(\varphi)/\partial
\varphi=0$ set at $\varphi=v\simeq246~\mathrm{GeV}$, where $m_{i}^{2}\equiv
m_{i}^{2}(\varphi)$ and $\dot{m}_{i}^{2}\equiv\partial m_{i}^{2}%
/\partial\varphi$. Then the Higgs mass at the one-loop level, which is nothing
but the second derivative of $V_{\mathrm{\mathrm{eff}}}^{T=0}(\varphi)$, is
given by%
\begin{equation}
m_{h}^{2}=\lambda_{1}v^{2}+\left.  \sum_{i}\frac{n_{i}}{32\pi^{2}%
}\Bigg[\Bigg(\ddot{m}_{i}^{2}m_{i}^{2}-\frac{\dot{m}_{i}^{2}m_{i}^{2}}%
{v}+\big(\dot{m}_{i}^{2}\big)^{2}\Bigg)\ln\frac{m_{i}^{2}}{\Lambda^{2}}%
-\ddot{m}_{i}^{2}m_{i}^{2}+\frac{\dot{m}_{i}^{2}m_{i}^{2}}{v}\Bigg]\right\vert
_{\varphi=v},\label{mh2}%
\end{equation}
where the first term is the familiar tree-level contribution, the second term
is the radiative one-loop correction, and $\ddot{m}_{i}^{2}\equiv\partial
^{2}m_{i}^{2}/\partial\varphi^{2}$. Accordingly, with $m_{h}^{2}$ being fixed
to its empirical value, as $\lambda_{1}$ is varied along with the other scalar
couplings it can be bigger or smaller than its tree-level value $\hat{\lambda
}_{1}=m_{h}^{2}/v^{2}\simeq0.258$, depending on the size and sign of the loop
contribution in eq (\ref{mh2}).

Incorporating eq.~(\ref{mh2}) into eq.~(\ref{l3h}), one then obtains
\begin{align}
\lambda_{hhh}  &  = \frac{3m_{h}^{2}}{v}+\frac{1}{32\pi^{2}}\sum
_{i=\mathrm{all}}n_{i} \Bigg\{ \Bigg[ \dddot{m}_{i}^{2} m_{i}^{2}%
+3\Bigg( \dot{m}_{i}^{2}-\frac{m_{i}^{2}}{v}\Bigg) \Bigg( \ddot{m}_{i}%
^{2}-\frac{\dot{m}_{i}^{2}}{v}\Bigg) \Bigg] \ln\frac{m_{i}^{2}}{\Lambda^{2}%
}\nonumber\\
&  ~~~~ + \frac{\left(  \dot{m}_{i}^{2}\right)  ^{3}}{m_{i}^{2}}-\dddot{m}%
_{i}^{2} m_{i}^{2}+\frac{3m_{i}^{2}}{v} \Bigg( \ddot{m}_{i}^{2}-\frac{\dot
{m}_{i}^{2}}{v}\Bigg) \Bigg\} \Bigg|_{\varphi=v} ,\label{lhhh}%
\end{align}
where $\dddot{m}_{i}^{2}\equiv\partial^{3}m_{i}^{2}/\partial\varphi^{3}$. Its
SM counterpart, $\lambda_{hhh}^{\mathrm{SM}}$, has the same formula, except
that in the sum $i$ runs over SM fields only.

\begin{figure}[t]
\includegraphics[width = 0.48\textwidth]{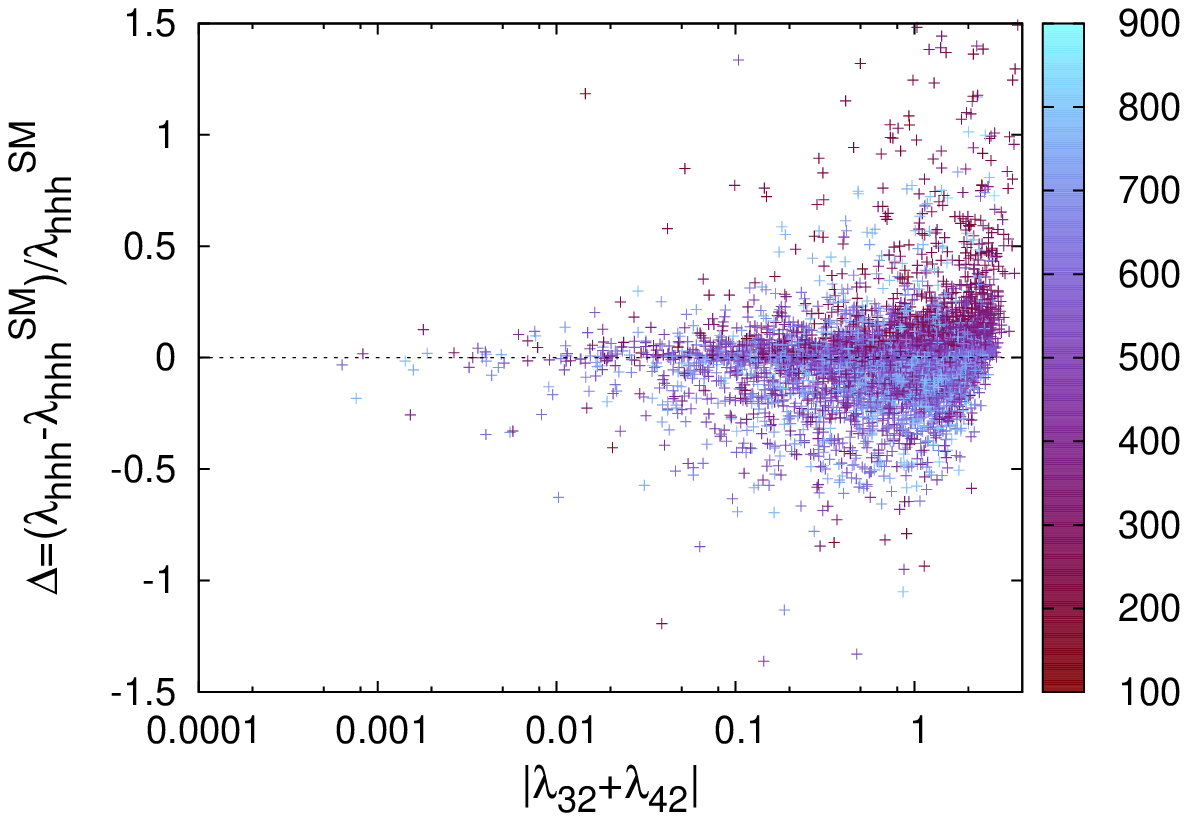}\includegraphics[width =
0.48\textwidth]{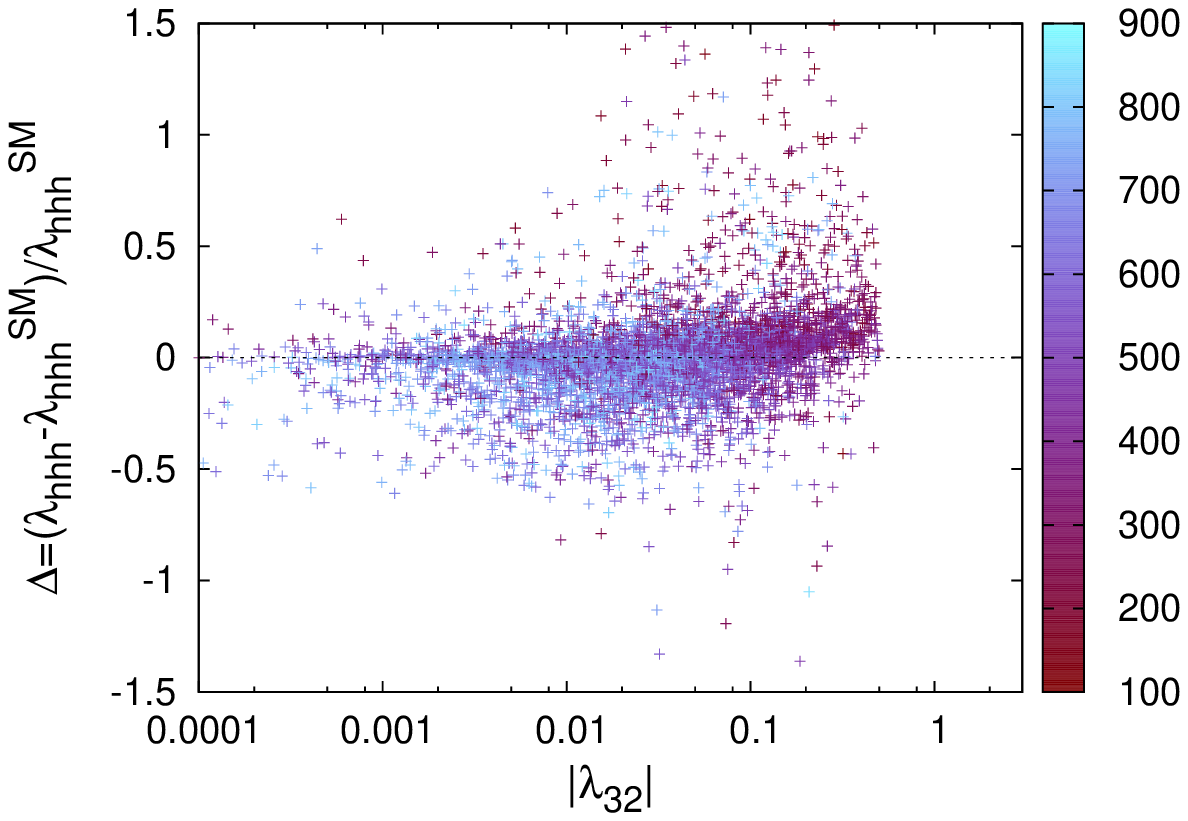}\vspace{-1ex}\caption{The changes of the Higgs
trilinear self-coupling relative to its SM value versus the absolute values of
the SM Higgs doublet couplings, $\lambda_{32}+\lambda_{42}$ and $\lambda_{32}%
$, to the heavy neutral (left) and the heavy charged (right) scalars,
respectively. On the palettes, we read the heavy neutral (left) and charged
(right) scalar masses in~GeV.}%
\label{Del}%
\end{figure}

According to eq. (\ref{lhhh}) and appendix \ref{masses}, the Higgs trilinear
self-coupling is a function of the couplings $\lambda_{3a}+\lambda_{4a}$ and
$\lambda_{3a}$ of the inert neutral and charged scalars, respectively, to the
SM Higgs doublet, i.e. through the field-dependent masses and their
derivatives. Since $\lambda_{3a,4a}$ are related to the scalars' physical
masses via eqs.~(\ref{mh}) and~(\ref{mx}), the Higgs trilinear coupling also
depends on them. To illustrate how the inert scalars' couplings and masses
affect $\lambda_{hhh}$, we define the relative change
\begin{equation}
\Delta= \frac{\lambda_{hhh}-\lambda_{hhh}^{\mathrm{SM}}}{\lambda
_{hhh}^{\mathrm{SM}}},
\end{equation}
with respect to the SM prediction. Then in figure \ref{Del} we graph $\Delta$
versus $|\lambda_{32}+\lambda_{42}|$ and $|\lambda_{32}|$, respectively, for
the 5000 benchmark points employed earlier. On the same plots we also show the
mass distributions of the inert neutral and charged scalars, respectively.

It is clear that in the presence of the inert doublets the trilinear Higgs
self-coupling can be enhanced or reduced by up to roughly 150\% relative to
the SM contribution to it. One realizes that, for either large or small
(charged and/or neutral) scalar masses and couplings to the SM Higgs doublet,
this enhancement or reduction of the trilinear coupling is the effect of the
superposition of different contributions which could be constructive or destructive.

\begin{figure}[t]
\begin{center}
\includegraphics[width = 0.48\textwidth]{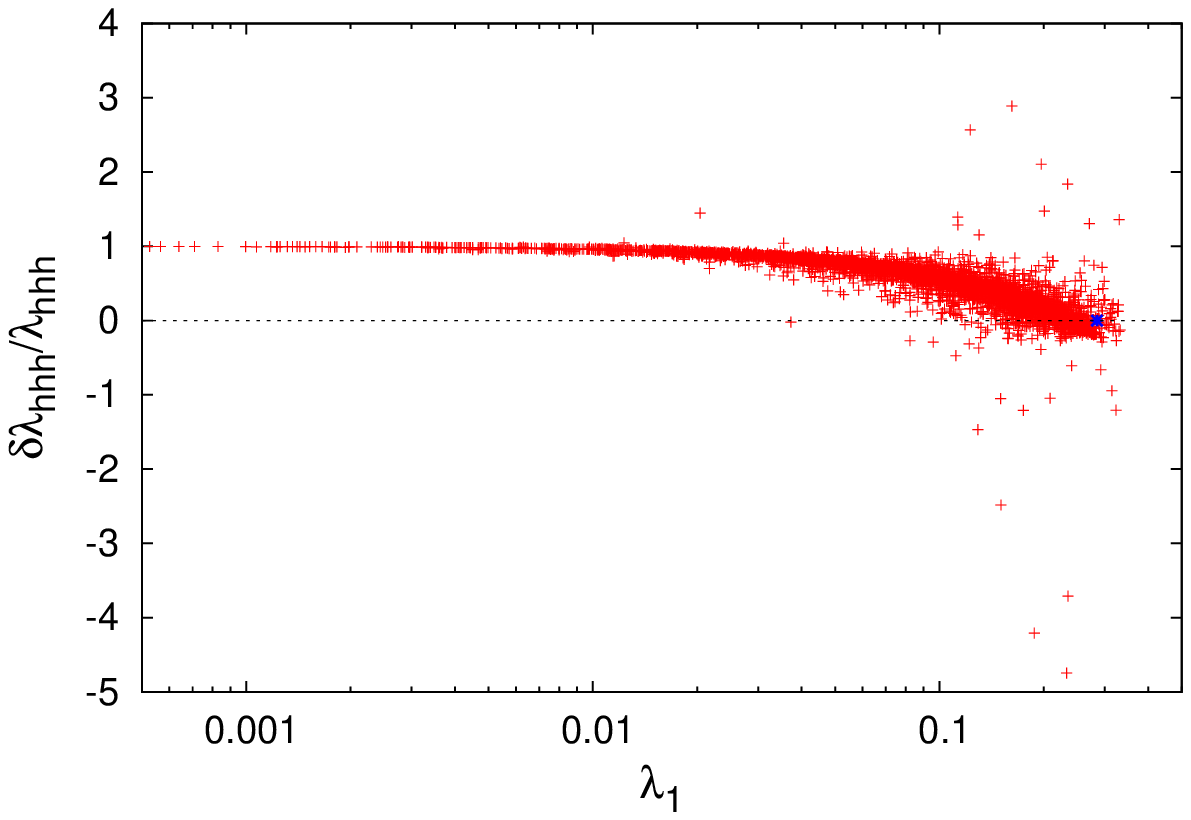}~\includegraphics[width =
0.48\textwidth]{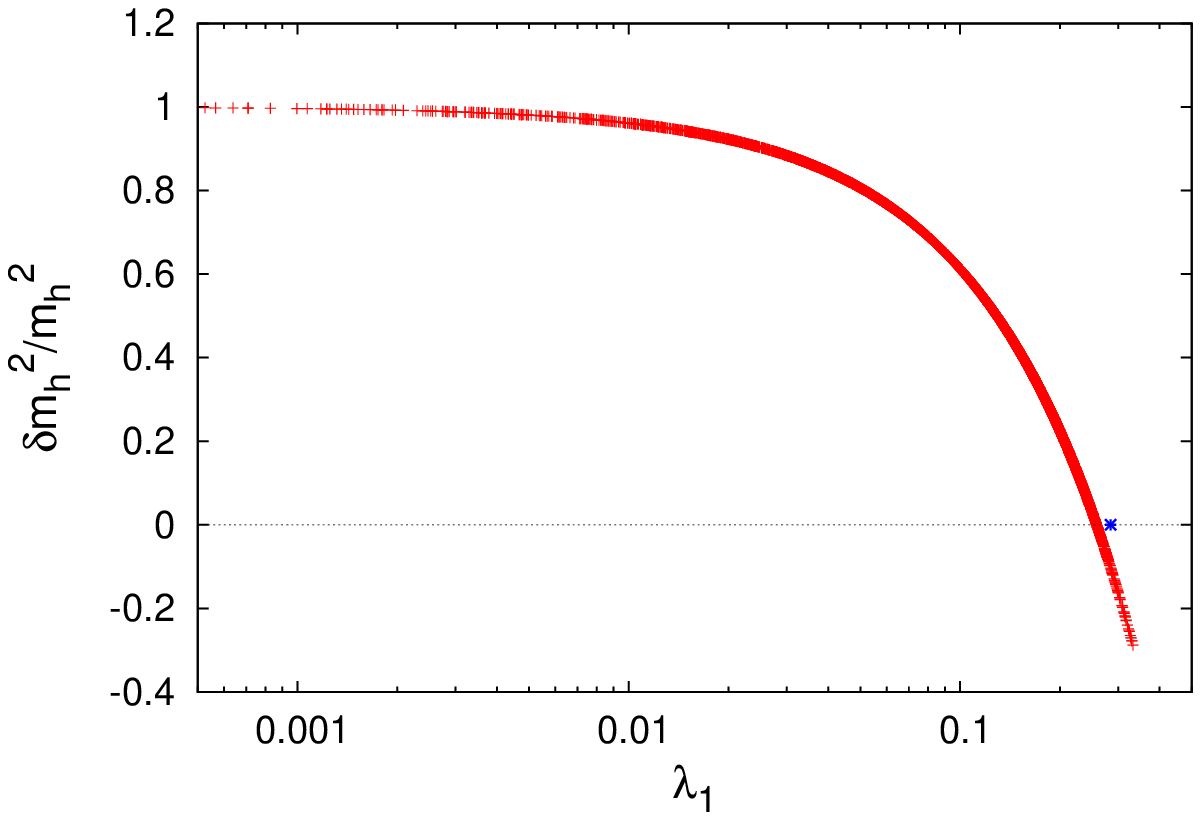} \includegraphics[width = 0.48\textwidth]{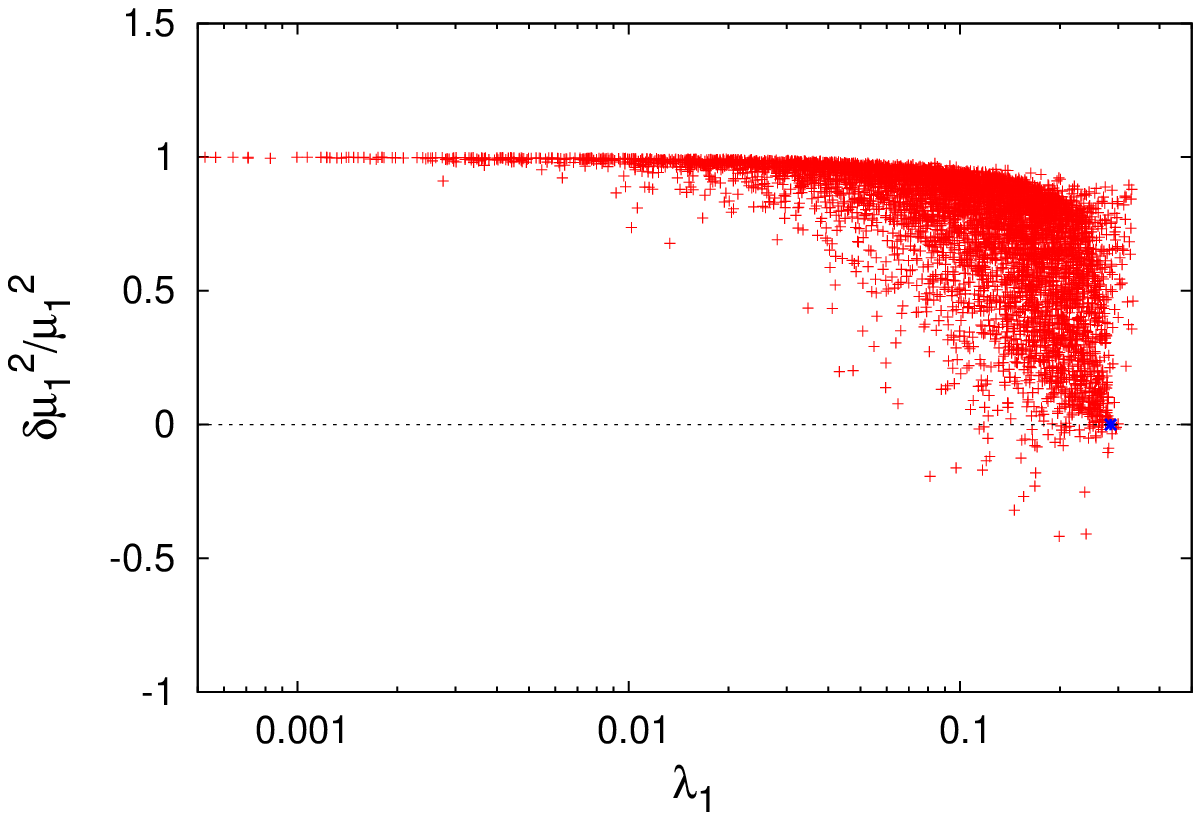}
\end{center}
\par
\vspace{-4ex}\caption{The relative changes of the trilinear Higgs
self-coupling (left), the Higgs mass (right) and the $\mu_{1}^{2}$ parameter
(bottom) due to loop corrections versus the Higgs quartic self-coupling
$\lambda_{1}$. The Higgs mass $m_{h}$ is fixed to 125.1\thinspace GeV. The
blue points represent the SM values.}%
\label{hhh}%
\end{figure}

The new scalars' impact can be further seen in figure~\ref{hhh}, which
illustrates their loop effects. Specifically, it displays the relative changes
of the trilinear Higgs coupling, the Higgs mass, and the parameter $\mu
_{1}^{2}$ due to radiative corrections versus the Higgs quartic self-coupling
$\lambda_{1}$, where
\begin{equation}
\delta\lambda_{hhh}=\lambda_{hhh}-3\lambda_{1}v,~~~~~\delta m_{h}^{2}%
=m_{h}^{2}-\lambda_{1}v^{2},
\end{equation}
$\delta\mu_{1}^{2}$ is defined in eq.~(\ref{dmu}), and $\mu_{1}^{2}=\hat{\mu
}_{1}^{2}+\delta\mu_{1}^{2}$.

We remark that the Higgs quartic self-coupling, which at tree level is defined
by the Higgs mass, can have a~wide range from about $10^{-4}$ to $0.5 $. This
is due to the fact that much of the Higgs mass arises radiatively, as the
right plot in figure~\ref{hhh} indicates. More precisely, $m_{h}$ can be fully
radiative for small $\lambda_{1}$ values or get a~negative radiative
correction for large $\lambda_{1}$ values, those greater than its tree-level
one,~$\hat{\lambda}_{1}$. One can see from the top-left and bottom plots in
the figure that similar remarks could be made concerning $\lambda_{hhh}$ and
$\mu_{1}^{2}$. In particular, each of these parameters may be fully radiative
for small $\lambda_{1}$ and also can receive radiative corrections which are negative.

\section{Electroweak Phase Transition\label{ewpt}}

It is well-known that one of the reasons why the SM fails to produce
successful baryogenesis \cite{EWB} is the fact that the EWPT is not strong and
consequently cannot suppress processes that violate the conservation of baryon
plus lepton numbers, $B$+$L$, in the broken phase \cite{Trodden:1998ym}. The
suppression of anomalous $B$+$L$-violating processes in the broken phase
happens if the criterion for strongly first-order EWPT
\cite{SFOPT,Shaposhnikov:1987pf},
\begin{equation}
v_{c}/T_{c}>1,\label{v/t}%
\end{equation}
is fulfilled, where $v_{c}$ is the Higgs VEV at the critical temperature
$T_{c}$ at which the effective potential exhibits two degenerate minima, one
at zero and the other at $v_{c}$. Both $T_{c}$ and $v_{c}$ are determined
using the full thermal effective potential \cite{Th,Weinberg:1974hy}
\begin{equation}
V_{\mathrm{eff}}(\varphi,T)=V_{\mathrm{eff}}^{T=0}(\varphi)+\frac{T^{4}}%
{2\pi^{2}}\sum_{i}n_{i}J_{\mathsf{B},\mathsf{F}}\big(m_{i}^{2}(\varphi
)/T^{2}\big)\label{VT}%
\end{equation}
at a finite temperature $T$, where
\begin{equation}
J_{\mathsf{B},\mathsf{F}}\left(  r\right)  =\int_{0}^{\infty}dx\;x^{2}%
\ln\Big[1\mp\exp\Big(\mbox{$-\sqrt{x^{2}+r}$}\Big)\Big],\label{JBF}%
\end{equation}
the upper (lower) sign referring to a boson (fermion). To $V_{\mathrm{eff}%
}(\varphi,T)$ one should add the so-called daisy (or ring) contribution
\cite{ring}
\begin{equation}
V_{\mathrm{ring}}(\varphi,T)=-\frac{T}{12\pi}\sum\limits_{i}n_{i}%
\bigl(\tilde{m}_{i}^{3}(\varphi,T)-m_{i}^{3}(\varphi)\bigr)\label{daisy}%
\end{equation}
which represents the leading term of higher-order loop corrections that may
play an important role during the EWPT dynamics. In $V_{\mathrm{ring}}%
(\varphi,T)$ the sum is over the scalar and longitudinal gauge degrees of
freedom, $\tilde{m}_{i}^{2}(\varphi,T)=m_{i}^{2}(\varphi)+\Pi_{i}(T)$ are
their thermal squared masses, and $\Pi_{i}(T)$ are the thermal parts of the
self energies, which are collected in appendix~\ref{masses}. To estimate
$V_{\mathrm{ring}}(\varphi,T)$, one performs the resummation of an infinite
class of infrared-divergent multiloops diagrams, known as ring diagrams, that
describes the dominant contribution of long distances and gives a significant
contribution when (almost) massless states appear in the system. In our case,
we will include this by following another approach. Rather than adding
$V_{\mathrm{ring}}(\varphi,T)$ to~$V_{\mathrm{eff}}(\varphi,T)$, we will
replace in eq.~(\ref{VT}) the field-dependent masses of the scalar and
longitudinal gauge degrees of freedom with their thermal masses $\tilde{m}%
_{i}(\varphi,T)$.

In the criterion for a strong first-order phase transition, eq.~(\ref{v/t}),
the critical temperature $T_{c}$ is the value at which the two minima of the
effective potential are degenerate,
\begin{equation}
\left.  \frac{\partial}{\partial\varphi}V_{\mathrm{eff}}(\varphi
,T_{c})\right\vert _{\varphi=v_{c}}=0,~~~~V_{\mathrm{eff}}(\varphi=v_{c}
,T_{c})=V_{\mathrm{eff}}(\varphi=0,T_{c}).\label{tc}%
\end{equation}
In the SM, this leads to a Higgs mass below 42 \textrm{GeV} \cite{mhbound},
since the ratio $v_{c}/T_{c}$ is inversely proportional to the Higgs quartic
coupling $\lambda_{1}$. The strength of the EWPT can be improved if new
bosonic degrees of freedom are
invoked~\cite{Chowdhury:2011ga,Ahriche:2010ny,Borah:2012pu,Gil:2012ya}, which
is the case we are investigating. It is clear from eq.~(\ref{mh2}) that for
large values of the couplings and/or masses of the extra scalars, the one-loop
corrections to the Higgs mass could be significant, which allows $\lambda_{1}
$ to be smaller and, therefore, fulfills the criterion in eq.~(\ref{v/t})
without conflicting with the recent Higgs mass measurements
\cite{mhx,CMS:2014ega}. Here, the relevant couplings are those of the Higgs
doublet to the charged scalars, $\lambda_{3a}$, and to the neutral ones,
$\lambda_{3a}+\lambda_{4a}$, in the limit $|\lambda_{5}|\ll|\lambda_{3a,4a}|$.
The situation may be compared to those in similar setups
\cite{KNT,hna,Ahriche:2012ei,kristian} where extra scalars can help bring
about a strongly first-order EWPT by (a)~relaxing the Higgs quartic coupling
$\lambda_{1}$ to as small as $\mathcal{O}(10^{-4})$ and (b)~enhancing the
value of the effective potential at the wrong vacuum at the critical
temperature without suppressing the ratio $v_{c}/T_{c}$, which relaxes the
severe bound on the mass of the SM Higgs.

The integral in eq.~(\ref{JBF}) is often approximated by a high temperature
expansion. However, in order to take into account the effect of all the (heavy
and light) degrees of freedom, we will evaluate them numerically.

\begin{figure}[t]
\vspace{1ex} \includegraphics[width =
0.5\textwidth]{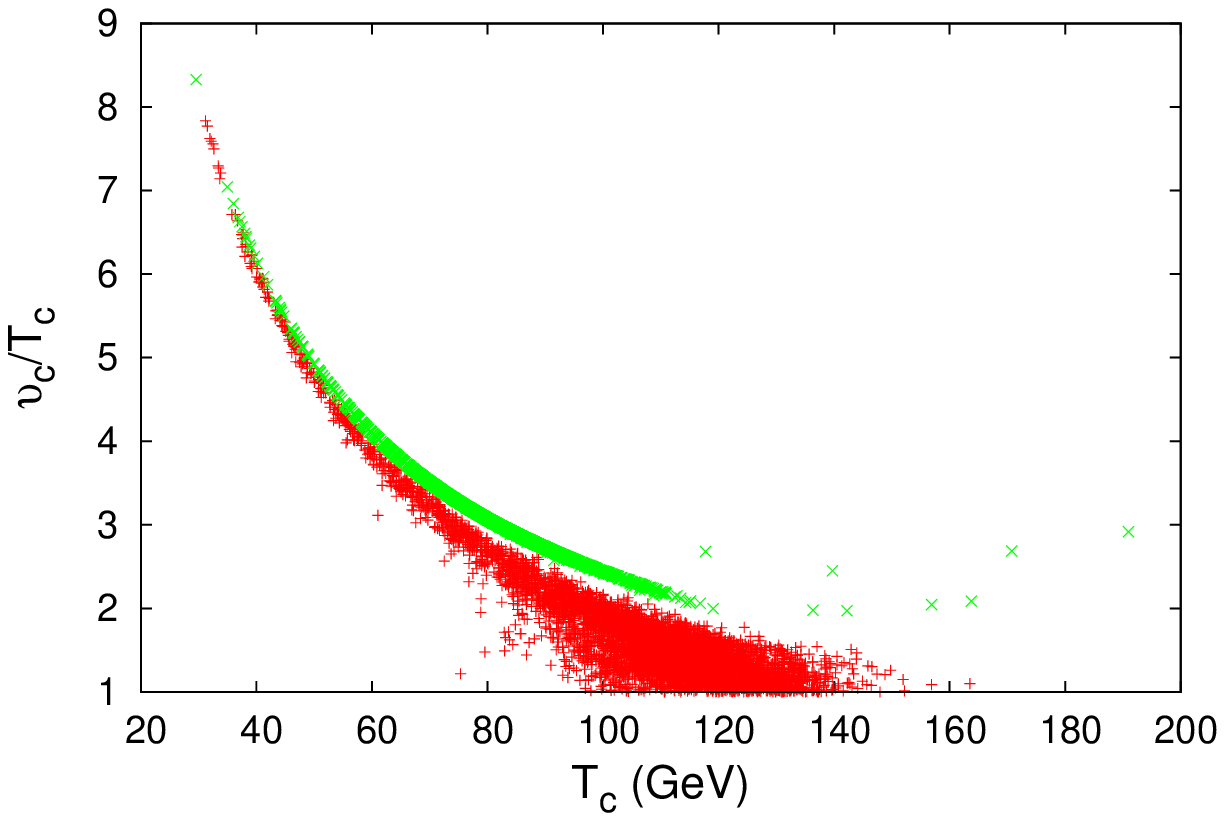}~\includegraphics[width =
0.5\textwidth]{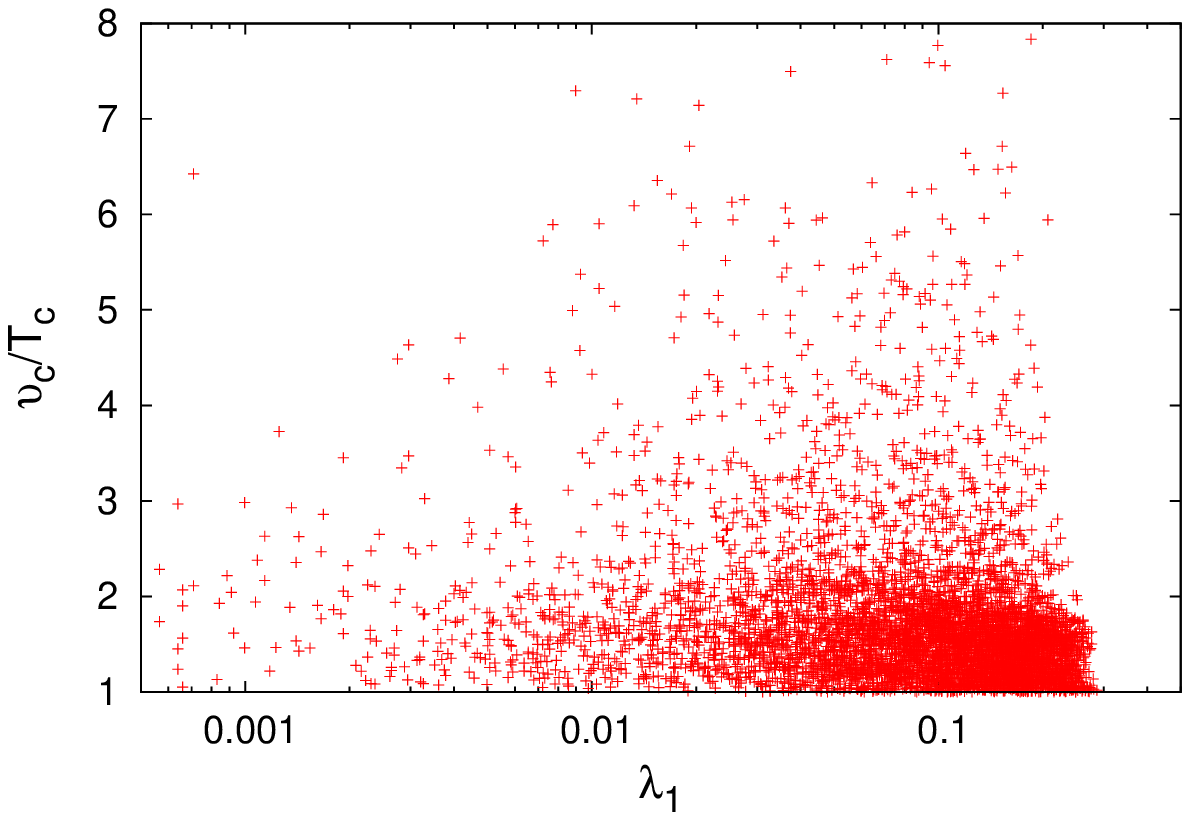}\vspace{-1ex}\caption{Left panel: $v_{c}/T_{c}$
versus $T_{c}$, estimated with (red) and without (green) the daisy
contribution. Right panel: $v_{c}/T_{c}$ versus the Higgs quartic
self-coupling $\lambda_{1}$, estimated by considering the daisy contribution.}%
\label{vctc}%
\end{figure}

With the same 5000 benchmark points used previously, in figure~\ref{vctc} we
present $v_{c}/T_{c}$ as a function of $T_{c}$ and of the Higgs quartic
self-coupling. It is obvious that the criterion for a strongly first-order
EWPT is easily satisfied for a large number of benchmarks. Moreover, we find
that the daisy contribution to the effective potential tends to weaken the
EWPT strength in this setup. One also notices that a strong EWPT can be
obtained for different values of the Higgs quartic self-coupling $\lambda_{1}
$, as shown in the right panel of figure~\ref{vctc}, even for values larger
than the tree-level one, $\hat{\lambda}_{1}$. This leads us to conclude that
the EWPT is always strongly first-order due the reason (b) mentioned above,
where the extra heavy scalars' existence makes the Higgs VEV slowly varying
with respect to temperature and the wrong vacuum value, i.e. $V_{\mathrm{eff}}
(\varphi=0,T)$, is evolving and increases with temperature.

We remark that due to the absence of a $CP$-violating phase in the potential
$\mathcal{V}$ an additional source of $CP$ violation has to be included in the
Lagrangian of the more complete theory for it to be realistic for baryogensis.
One possibility is to introduce dimension-six operators which couple the inert
scalars to the top-quark mass and are suppressed by a new-physics scale that
can be well above one \textrm{TeV}, in analogy to a~scenario of electroweak
baryogenesis from a singlet scalar \cite{Cline}.

\section{Discussion \& Conclusion\label{concl}}

\begin{figure}[h]
\includegraphics[width = 0.48\textwidth]{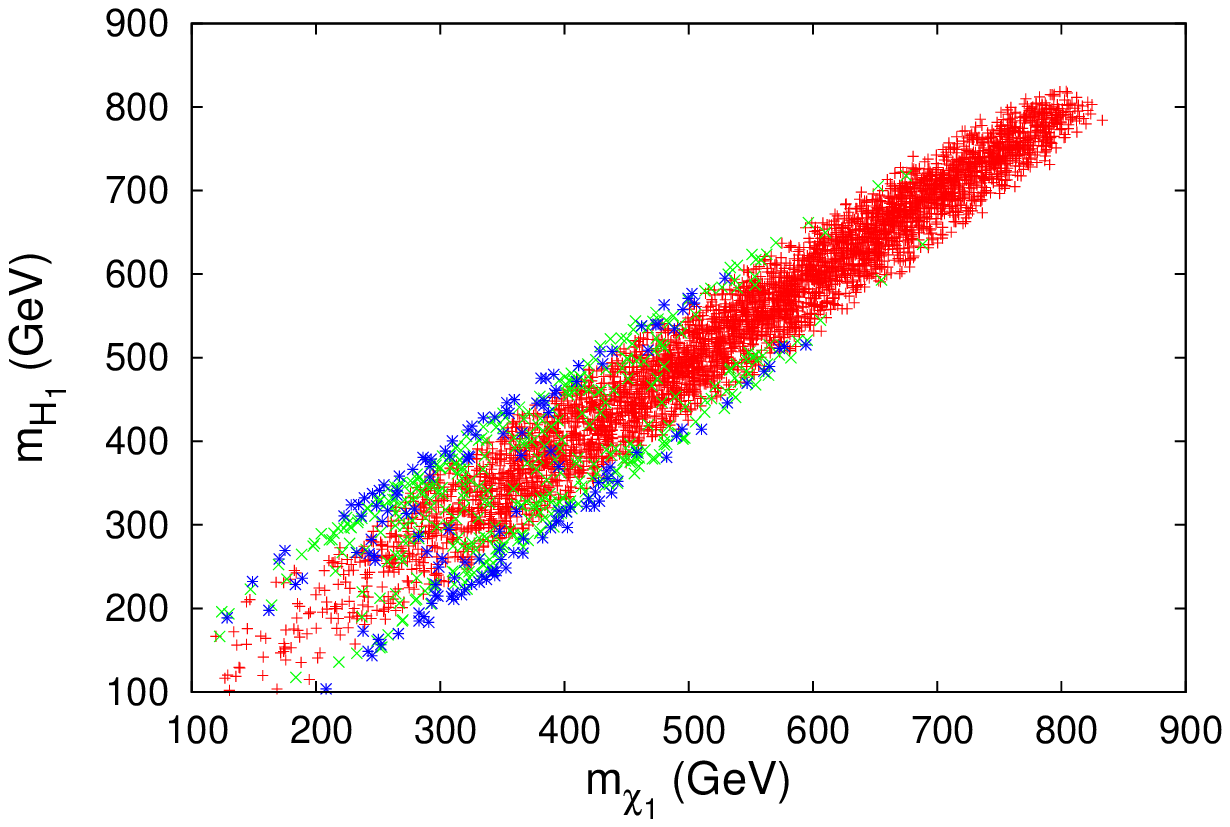}~\includegraphics[width =
0.48\textwidth]{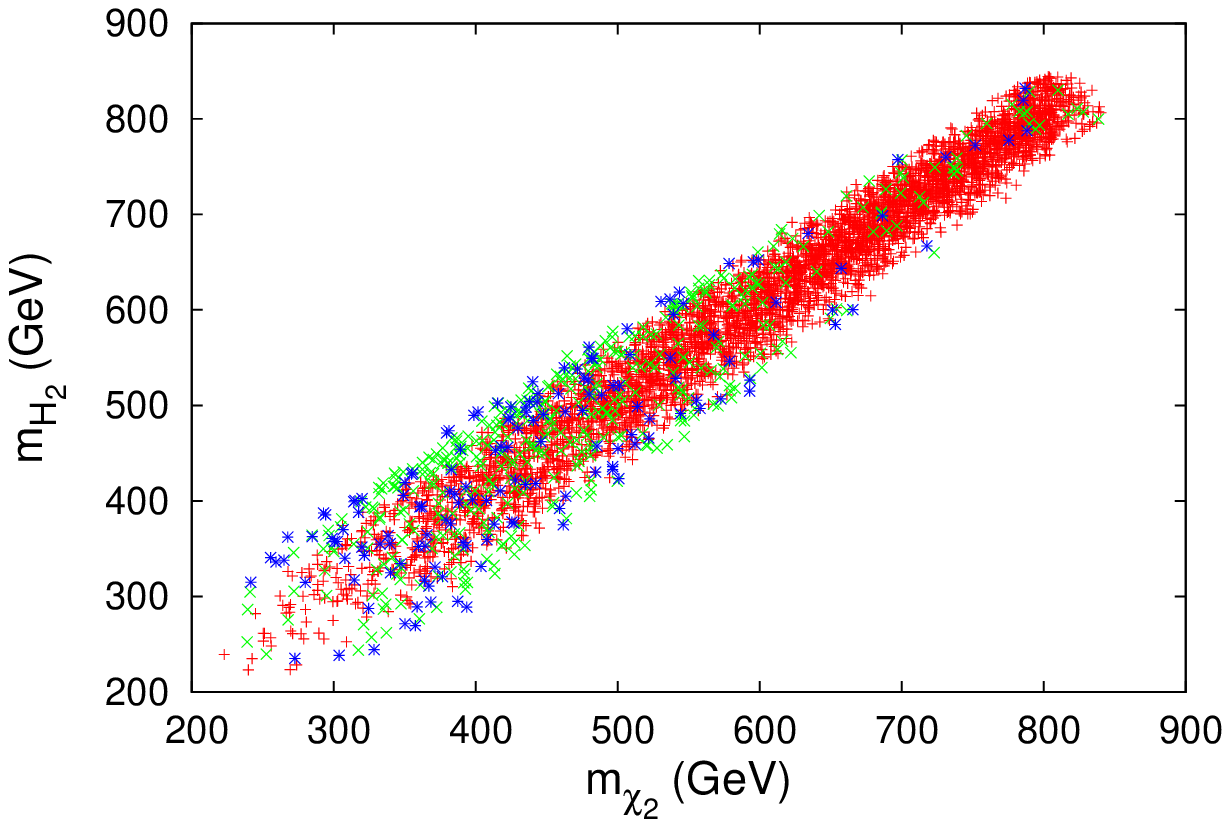}\newline%
\includegraphics[width = 0.48\textwidth]{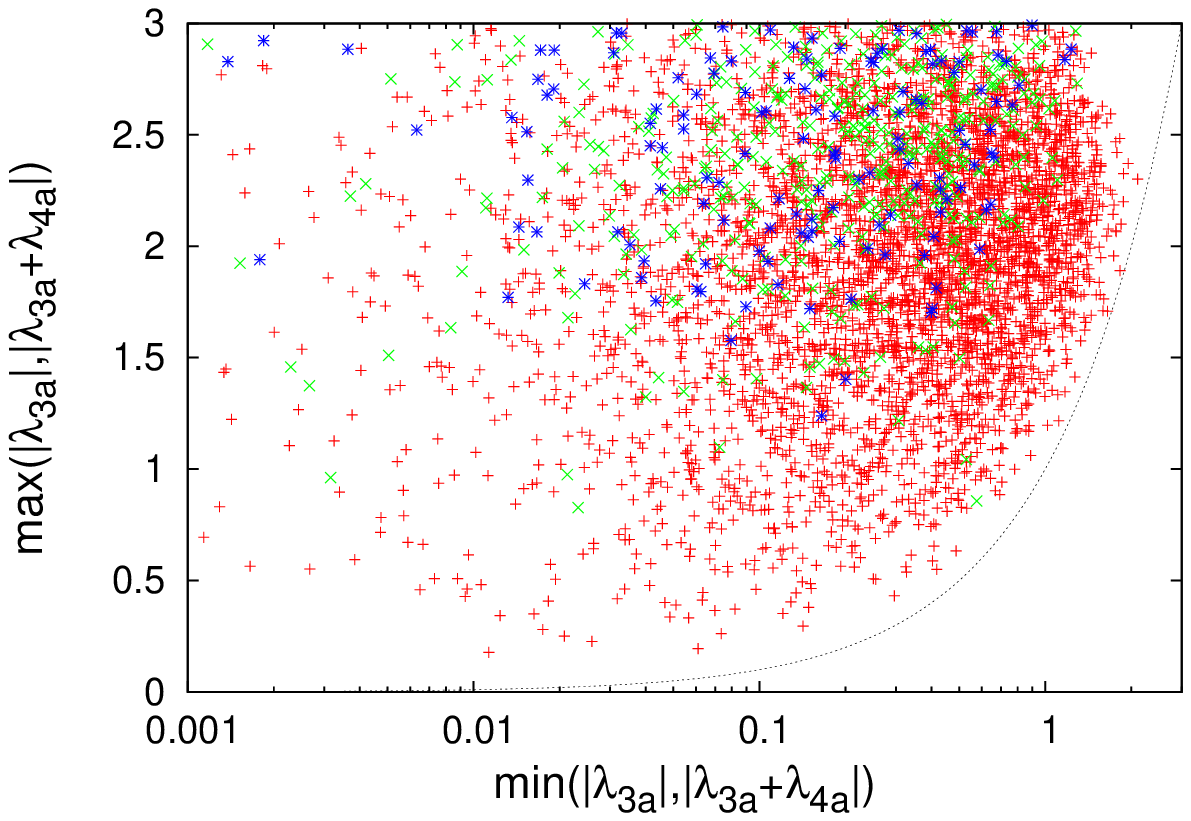}~\includegraphics[width =
0.48\textwidth]{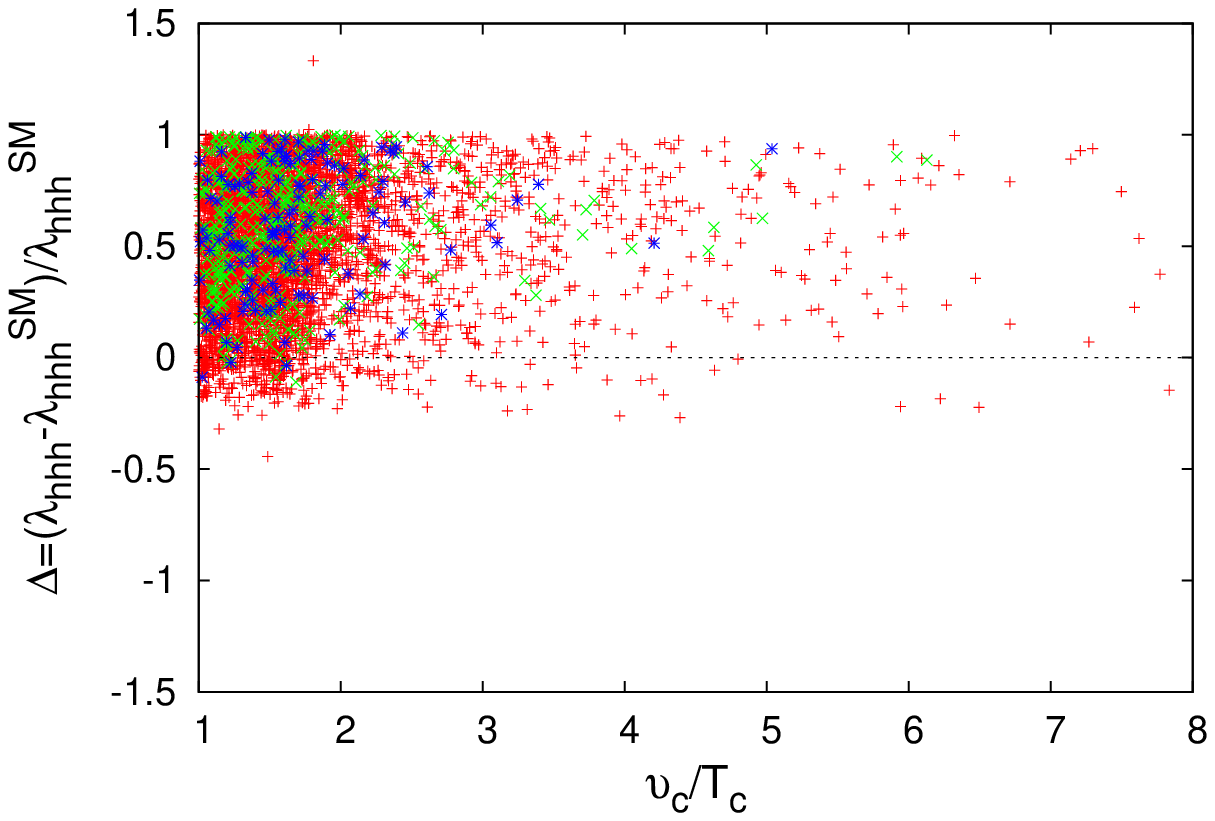}\hspace{\fill} \vspace{-1em}\caption{Top panels:
the distribution of masses of the extra charged and neutral scalars under the
assumptions described in the text. Bottom left panel: the strength
distribution of the different quartic self-coupling of the Higgs doublet to
the extra charged and neutral scalars; the dashed line represents the case
where the couplings are equal in magnitude, i.e., $\mathrm{Min}(|\lambda
_{3a}|,|\lambda_{3a}+\lambda_{4a}|)=\mathrm{Max} (|\lambda_{3a}|,|\lambda
_{3a}+\lambda_{4a}|)$. Bottom right panel: the EWPT strength versus the
relative enhancement of the Higgs trilinear self-coupling. The red, green, and
blue points correspond to those inside the 99\%, 95\%, and 68\% CL ellipsoids
on the $(\Delta S,\Delta T)$ plane in figure~\ref{ob}.}%
\label{pr}%
\end{figure}

According to the analysis carried out in previous sections, the extra scalars
can have important effects on the Higgs phenomenology and the electroweak
phase transition if these particles are relatively light and the couplings to
the SM Higgs doublet are large ($\lambda_{3a}$ for charged scalars and
$\lambda_{3a}+\lambda_{4a}$ for neutral ones). Therefore, from the 5000
benchmark points used previously, we extract those that simultaneously satisfy
(i)~the constraint from the measurements on the Higgs decay mode
$h\rightarrow\gamma\gamma$, namely $0.9<\mathcal{R}_{\gamma\gamma}<1.37$,
(ii)~the electroweak precision tests, i.e. all the points inside the three
ellipsoids in figure~\ref{ob}, and (iii)~the criterion $v_{c}/T_{c}>1$ for
strongly first-order~EWPT. As mentioned in section~\ref{ewpconstraints}, the
Higgs decay channel into a pair of inert scalars is closed for all the viable
benchmarks and hence its experimental bound is not relevant. Here, we divide
the points fulfilling the conditions (i,ii,iii) into three sets according to
the ellipsoid to which they belong on the $(\Delta S,\Delta T)$ plane. The
results are displayed in figure~\ref{pr}.

From the top panels in figure \ref{pr}, one can see that the extra scalar
masses do not exceed 900 \textrm{GeV} according to our parameter choices in
eq.~(\ref{PAR}). The charged scalars could be light up to the LEP II bound
(100 \textrm{GeV}), while the neutral scalars, which were supposed to be less
constrained before, are now not allowed to be less than 120~\textrm{GeV} due
to the electroweak precision tests in this model. From the bottom left panel,
it is evident that the couplings of the Higgs doublet to the charged scalars,
{$\lambda_{3a}$}, and to the neutral ones, $\lambda_{3a}+\lambda_{4a}$, could
be both larger than 1 or smaller than $0.5$. They could vary also within the
whole considered range [0:3], or they could be almost equal in absolute values
(i.e. close to the dashed curve). The bottom right plot in this figure reveal
that, while strongly first-order EWPT occurs for all of the viable benchmark
points, only for some of them there is a positive correlation between the EWPT
strength and substantial enhancement of the Higgs trilinear self-coupling
relative to the SM prediction as shown in ref.~\cite{KOS}.

In conclusion, we have considered a scenario beyond the SM involving three
scalar weak doublets and investigated a number of implications of the case
where two of the doublets are inert and charged under a dark Abelian gauge
symmetry. We looked at the effects of the new scalars on oblique electroweak
parameters, the Higgs decay modes $h\rightarrow\gamma\gamma,\gamma Z$, and its
trilinear coupling. We also examined how the inert scalars can induce strongly
first-order EWPT. Taking into account various theoretical and experimental
constraints, we demonstrated that the viable parameter space can all
accommodate strongly first-order EWPT and contains regions in which the Higgs
trilinear self-coupling is enhanced/reduced by up to 150\% compared to its SM
value. Future experiments with sufficient precision can test the new scalars'
effects that we have obtained on the Higgs decays $h\rightarrow\gamma
\gamma,\gamma Z$ and trilinear coupling.

\acknowledgments

We would like to thank Masaya Kohda for helpful discussions. A.A. is supported
by the Algerian Ministry of Higher Education and Scientific Research under the
CNEPRU Project No. \textit{D01720130042}. The work of G.F. and J.T. was
supported in part by the research grant NTU-ERP-102R7701, the MOE Academic
Excellence Program (Grant No. 102R891505), and the National Center for
Theoretical Sciences of Taiwan.

\appendix

\section{Vacuum stability conditions\label{app:vacstab}}

We can rewrite the doublets $\Phi$ and $\eta_{1,2}$ and their products
according to
\begin{align}
&  \Phi= f_{ }\hat{\Phi} ,~~~~~\hat{\Phi}^{\dagger}\hat{\Phi} = 1
,~~~~~~~\eta_{a} = e_{a}\hat{\eta}_{a} ,~~~~~\hat{\eta}_{a}^{\dagger}\hat
{\eta}_{a} = 1\;,~~~~~~~f,e_{a} > 0 ,\nonumber\\
&  \hat{\Phi}^{\dagger}\hat{\eta}_{a} \hat{\eta}_{a}^{\dagger}\hat{\Phi} =
\rho_{a} ,~~~~~~~\hat{\eta}_{1}^{\dagger}\hat{\eta}_{2} \hat{\eta}%
_{2}^{\dagger}\hat{\eta}_{1} = \rho^{\prime} ,~~~~~~~0 \leq\rho_{a}%
,\rho^{\prime}\leq1 .~~~~~~~
\end{align}
Assuming that $\lambda_{5}$ in eq.~(\ref{potential}) is negligible compared to
the other $\lambda$'s, we can then express the part of $\mathcal{V}$ that is
quartic in the doublets approximately as
\begin{align}
\mathcal{V}_{4}  &  = \tfrac{1}{2}\lambda_{1 }f^{4}+\tfrac{1}{2}\lambda_{21
}e_{1}^{4}+\tfrac{1}{2}\lambda_{22 }e_{2}^{4} + \lambda_{31 }f^{2}e_{1}%
^{2}+\lambda_{32 }f^{2}e_{2}^{2}\nonumber\label{V4}\\
&  ~~~ +\;\lambda_{41 }f^{2}e_{1 }^{2}\rho_{1}+\lambda_{42 }f^{2}e_{2}^{2
}\rho_{2} + \lambda_{6 }e_{1}^{2}e_{2}^{2}+\lambda_{7 }e_{1}^{2}e_{2}^{2 }%
\rho^{\prime}\nonumber\\
&  = \tfrac{1}{2}\left(
\begin{array}
[c]{ccc}%
f^{2} & e_{1}^{2} & e_{2}^{2}%
\end{array}
\right)  \tilde\lambda\left(
\begin{array}
[c]{c}%
f^{2}\vspace{1pt}\\
e_{1}^{2}\vspace{1pt}\\
e_{2}^{2}%
\end{array}
\right)  ,
\end{align}
where
\begin{equation}
\tilde\lambda=\left(
\begin{array}
[c]{ccccc}%
\lambda_{1} &  & \lambda_{31}+\rho_{1}\lambda_{41} &  & \lambda_{32}+\rho
_{2}^{\;\;}\lambda_{42}\vspace{2pt}\\
\lambda_{31}+\rho_{1}\lambda_{41} &  & \lambda_{21} &  & \lambda_{6}%
+\rho^{\prime}\lambda_{7}\vspace{2pt}\\
\lambda_{32}+\rho_{2}^{\;\;}\lambda_{42} &  & \lambda_{6}+\rho^{\prime}%
\lambda_{7} &  & \lambda_{22}%
\end{array}
\right)  .
\end{equation}
To ensure the stability of the vacuum, we need to derive relations among the
$\lambda$'s in~$\mathcal{V}_{4}$, which dominates $\mathcal{V}$ at large
fields, such that the minimum of $\mathcal{V}_{4}$ remains positive. This can
be achieved using copositivity criteria \cite{Kannike:2012pe}, which in this
case are applied to the minimum of $\tilde\lambda$. Since $\lambda_{4a,7}$ can
be positive, zero, or negative and $0\leq\rho_{a},\rho^{\prime}\leq1$,\ we
have
\begin{equation}
\tilde\lambda_{\mathrm{min}} = \left(
\begin{array}
[c]{ccccc}%
\lambda_{1} &  & \lambda_{31}+\mathrm{Min}\left(  0,\lambda_{41}\right)  &  &
\lambda_{32}+\mathrm{Min}\left(  0,\lambda_{42}\right) \\
\lambda_{31}+\mathrm{Min}\left(  0,\lambda_{41}\right)  &  & \lambda_{21} &  &
\lambda_{6}+\mathrm{Min}\left(  0,\lambda_{7}\right) \\
\lambda_{32}+\mathrm{Min}\left(  0,\lambda_{42}\right)  &  & \lambda
_{6}+\mathrm{Min}\left(  0,\lambda_{7}\right)  &  & \lambda_{22}%
\end{array}
\right)  .
\end{equation}
From the criteria for strictly copositive 3$\times$3 matrices
\cite{copositivity1,copositivity2,copositivity3} then follow the conditions in
eq.~(\ref{vacs}).

\section{Interaction terms for $\mathcal{S}_{a}$ and $\mathcal{P}_{a}%
$\label{interactions}}

The interaction terms of $\chi_{1,2}$ in eqs.~(\ref{Lint}) and (\ref{h2xx})
can be rewritten in terms of the real and imaginary components defined in
eq.~(\ref{SaPa}). Thus
\begin{align}
\mathcal{L}  &  \supset\frac{g}{2c_{\mathrm{w}}}\Big[c_{2\theta}%
\Big(\mathcal{P}_{1}%
\raisebox{-0.7pt}{\small$\stackrel{\scriptscriptstyle\leftrightarrow}{\partial}$}{}%
^{\mu}\mathcal{S}_{1}-\mathcal{P}_{2}%
\raisebox{-0.7pt}{\small$\stackrel{\scriptscriptstyle\leftrightarrow}{\partial}$}{}%
^{\mu}\mathcal{S}_{2}\Big)+s_{2\theta}\Big(\mathcal{P}_{1}%
\raisebox{-0.7pt}{\small$\stackrel{\scriptscriptstyle\leftrightarrow}{\partial}$}{}%
^{\mu}\mathcal{S}_{2}+\mathcal{P}_{2}%
\raisebox{-0.7pt}{\small$\stackrel{\scriptscriptstyle\leftrightarrow}{\partial}$}{}%
^{\mu}\mathcal{S}_{1}\Big)\Big]Z_{\mu}\nonumber\\
&  ~~~~+\frac{ig}{2}\Big\{\Big[c_{\theta}H_{1}^{+}%
\raisebox{-0.7pt}{\small$\stackrel{\scriptscriptstyle\leftrightarrow}{\partial}$}{}%
^{\mu}(\mathcal{S}_{1}-i\mathcal{P}_{1})+c_{\theta}H_{2}^{+}%
\raisebox{-0.7pt}{\small$\stackrel{\scriptscriptstyle\leftrightarrow}{\partial}$}{}%
^{\mu}(\mathcal{S}_{2}+i\mathcal{P}_{2})\nonumber\\
&  \hspace{5em}+s_{\theta}H_{1}^{+}%
\raisebox{-0.7pt}{\small$\stackrel{\scriptscriptstyle\leftrightarrow}{\partial}$}{}%
^{\mu}(\mathcal{S}_{2}-i\mathcal{P}_{2})-s_{\theta}H_{2}^{+}%
\raisebox{-0.7pt}{\small$\stackrel{\scriptscriptstyle\leftrightarrow}{\partial}$}{}%
^{\mu}(\mathcal{S}_{1}+i\mathcal{P}_{1})\Big]W_{\mu}^{-}-\mathrm{H.c.}%
\Big\}\nonumber\\
&  ~~~~+\frac{g^{2}}{4}\big(\mathcal{S}_{1}^{2}+\mathcal{P}_{1}^{2}%
+\mathcal{S}_{2}^{2}+\mathcal{P}_{2}^{2}\big)\left(  \frac{Z^{2}%
}{2c_{\mathrm{w}}^{2}}+W^{+\mu}W_{\mu}^{-}\right) \nonumber\\
&  ~~~~+\frac{h}{v}\bigl[\bigl(c_{\theta}^{2}\mu_{21}^{2}+s_{\theta}^{2}%
\mu_{22}^{2}-m_{\chi_{1}}^{2}\bigr)\bigl(\mathcal{S}_{1}^{2}+\mathcal{P}%
_{1}^{2}\bigr)+\bigl(c_{\theta}^{2}\mu_{22}^{2}+s_{\theta}^{2}\mu_{21}%
^{2}-m_{\chi_{2}}^{2}\bigr)\bigl(\mathcal{S}_{2}^{2}+\mathcal{P}_{2}%
^{2}\bigr)\nonumber\\
&  \hspace{4em}+s_{2\theta}\bigl(\mu_{21}^{2}-\mu_{22}^{2}%
\bigr)\bigl(\mathcal{S}_{1}\mathcal{S}_{2}+\mathcal{P}_{1}\mathcal{P}%
_{2}\bigr)\bigr]\;.
\end{align}

\section{Field-Dependent and Thermal Masses\label{masses}}

To estimate the Higgs effective potential, one needs the field-dependent
squared masses $m_{i}^{2}(\varphi)$ of all the contributing particles. One
also requires the first, second, and third derivatives of $m_{i}^{2}(\varphi)
$ to determine the counterterm $\delta\mu_{1}^{2}$ in eq.~(\ref{dmu}), the
one-loop correction to the Higgs mass, and the enhancement of the Higgs
trilinear self-coupling.

The field-dependent masses of the electroweak gauge bosons and top quark have
their SM values. For the other particles, we have the thermal masses
$\tilde{m}_{i}\equiv\tilde{m}_{i}(\varphi,T)$ which are given by
\begin{align}
\tilde{m}_{\mathcal{G}}^{2} &  =\mu_{1}^{2}+\tfrac{1}{2}\lambda_{1}\varphi
^{2}+\Pi_{\Phi},\hspace{11ex}\tilde{m}_{h}^{2}=\mu_{1}^{2}+\tfrac{3}{2}%
\lambda_{1}\varphi^{2}+\Pi_{\Phi},\nonumber\label{m(phi)}\\
\tilde{m}_{H_{a}^{\pm}}^{2} &  =\mu_{2a}^{2}+\tfrac{1}{2}\lambda_{{3a}}%
\varphi^{2}+\Pi_{\eta_{a}},\hspace{6ex}\tilde{m}_{\chi_{1,2}}^{2}=\tfrac{1}%
{2}\Big(C_{1}+C_{2}\mp\sqrt{R}\Big),\nonumber\\
C_{a} &  =\mu_{2a}^{2}+\tfrac{1}{2}(\lambda_{3a}+\lambda_{4a})\varphi^{2}%
+\Pi_{\eta_{a}},~~~R=\left(  C_{1}-C_{2}\right)  ^{2}+4c^{2},~~c=\tfrac{1}%
{4}\left\vert \lambda_{5}\right\vert \varphi^{2},~
\end{align}
and related to $m_{i}(\varphi)$ by $\tilde{m}_{i}^{2}(\varphi,T)=m_{i}%
^{2}(\varphi)+\Pi_{i}$, where $\Pi_{i}\equiv\Pi_{i}(T)$ denote the thermal
parts of the self energies and $\Pi_{\Phi,\eta_{a}}$ are listed below. Hence
the Goldstone bosons ($\mathcal{G}$) and the Higgs boson also have the same
field-dependent masses as their respective counterparts in the SM. We note
that the inert $CP$-even and $CP$-odd neutral scalars mix, leading to
equal-mass eigenstates, according to eq.~(\ref{mix}).

It is simple to get the first, second, and third derivatives of $m_{i}%
^{2}(\varphi)$ from eq.~(\ref{m(phi)}). For completeness, here we supply them
explicitly:%
\begin{align}
\dot{m}_{\mathcal{G}}^{2}(\varphi) &  =\lambda_{1}\varphi,~~~\dot{m}_{h}%
^{2}(\varphi)=3\lambda_{1}\varphi,~~~\dot{m}_{H_{a}^{\pm}}^{2}(\varphi
)=\lambda_{3a}\varphi,\nonumber\\
\dot{m}_{\chi_{1,2}}^{2}(\varphi) &  =\frac{1}{2}\left(  \dot{C}_{1}+\dot
{C}_{2}\mp\frac{\dot{R}}{2\sqrt{R}}\right)  ,\nonumber\\
\dot{C}_{a} &  =\left(  \lambda_{{3a}}+\lambda_{{4a}}\right)  \varphi
,~~~\dot{c}=\tfrac{1}{2}\left\vert \lambda_{{5}}\right\vert \varphi
,\nonumber\\
\dot{R} &  =2\left(  \dot{C}_{1}-\dot{C}_{2}\right)  \left(  C_{1}%
-C_{2}\right)  +8\dot{c}c,
\end{align}%
\begin{align}
\ddot{m}_{\mathcal{G}}^{2}(\varphi) &  =\lambda_{1},~~~\ddot{m}_{h}%
^{2}(\varphi)=3\lambda_{1},~~~\ddot{m}_{H_{a}^{\pm}}^{2}(\varphi
)=\lambda_{{3a}},\nonumber\\
\ddot{m}_{\chi_{1,2}}^{2}(\varphi) &  =\frac{1}{2}\left(  \ddot{C}_{1}%
+\ddot{C}_{2}\mp\frac{\ddot{R}}{2\sqrt{R}}\pm\frac{\dot{R}^{2}}{4\sqrt{R^{3}}%
}\right)  ,\nonumber\\
\ddot{C}_{a} &  =\lambda_{{3a}}+\lambda_{{4a}},~~~\ddot{c}=\tfrac{1}%
{2}\left\vert \lambda_{{5}}\right\vert ,\nonumber\\
\ddot{R} &  =2\left(  \dot{C}_{1}-\dot{C}_{2}\right)  ^{2}+2\left(  \ddot
{C}_{1}-\ddot{C}_{2}\right)  \left(  C_{1}-C_{2}\right)  +8\dot{c}^{2}%
+8\ddot{c}c,
\end{align}%
\begin{align}
\dddot{m}_{\mathcal{G}}^{2}(\varphi) &  =\dddot{m}_{h}^{2}(\varphi)=\dddot
{m}_{H_{a}^{\pm}}^{2}(\varphi)=\dddot{C}_{a}=\dddot{c}=0,\nonumber\\
\dddot{m}_{\chi_{1,2}}^{2}(\varphi) &  =\mp\frac{1}{4\sqrt{R}}\left(
\dddot{R}-\frac{3\dot{R}\ddot{R}}{2R}+\frac{3\dot{R}^{3}}{4R^{2}}\right)
,\nonumber\\
\dddot{R} &  =6\left(  \ddot{C}_{1}-\ddot{C}_{2}\right)  \left(  \dot{C}%
_{1}-\dot{C}_{2}\right)  +24\ddot{c}\dot{c}.
\end{align}

Finally, we write down the thermal parts $\Pi_{i}$ of the pertinent
self-energies. For the scalar and electroweak bosons \cite{kapusta}
\begin{align}
\Pi_{\Phi}  &  =\left(  6\lambda_{{1}}+\frac{9}{4}{g}^{2}+\frac{3}{4}g_{Y}%
^{2}+3y_{t}^{2}+4\lambda_{{31}}+2\lambda_{{41}}+4\lambda_{{32}}+2\lambda
_{{42}}\right)  \frac{T^{2}}{12},\nonumber\\
\Pi_{\eta_{a}}  &  =\left(  \frac{9}{4}{g}^{2}+\frac{3}{4}g_{Y}^{2}%
+4\lambda_{{3a}} +2\lambda_{{4a}}+6\lambda_{{2a}}+4\lambda_{{6}}+2\lambda
_{{7}} +\frac{3}{4}\mathcal{Q}_{\eta_{a}}^{2}g_{D}^{2}\right)  \frac{T^{2}%
}{12},\nonumber\\
\Pi_{W}  &  =\frac{17}{6}g^{2}T^{2},~~~\Pi_{B}=\frac{11}{16}g_{Y}^{2}T^{2},
\end{align}
where $y_{t}$ denotes the top-quark Yukawa coupling and $\mathcal{Q}_{\eta
_{a}}=\mathcal{Q}_{C}\eta_{a}$ is the charge of the inert doublet $\eta_{a}$
under U(1)$_{D}$. Numerically, since $g_{D}$ is unknown, for definiteness we
set $g_{D}=g_{Y}$.

\end{document}